\documentclass[twocolumn,showpacs,preprintnumbers,amsmath,amssymb,floatfix,superscriptaddress]{revtex4-1}

 \usepackage[utf8]{inputenc}

 \usepackage[colorlinks,
	linkcolor=blue,
	citecolor=blue,
	urlcolor=blue]{hyperref}

\usepackage[dvips]{graphicx}
\usepackage{dcolumn}
\usepackage{bm}
\usepackage{bbm}
\usepackage{dsfont}
\usepackage{amsmath,amsthm,amssymb}
\usepackage{graphicx}
\usepackage{epstopdf}
\usepackage{psfrag}
\usepackage{mathrsfs}
\usepackage{color}

\def\beq{\begin{equation}}
\def\eeq{\end{equation}}

\newcommand{\mH}[0]{\mathcal{H}}
\newcommand{\mC}[0]{\mathcal{C}}

\newcommand{\mD}[0]{\mathcal{D}}
\newcommand{\mQ}[0]{\mathcal{Q}}
\newcommand{\md}[0]{\nu}

\newcommand{\ev}[1]{\langle #1 \rangle}

\newcommand{\dx}[0]{\delta x}
\newcommand{\dy}[0]{\delta y}
\newcommand{\dz}[0]{\delta z}
\newcommand{\ba}[0]{\mathbf a}

\newcommand{\bp}[0]{\mathbf p}
\newcommand{\bq}[0]{\mathbf q}
\newcommand{\bP}[0]{\mathbf P}
\newcommand{\bb}[0]{\mathbf b}
\newcommand{\be}[0]{\mathbf e}
\newcommand{\bu}[0]{\mathbf u}
\newcommand{\bv}[0]{\mathbf v}
\newcommand{\bw}[0]{\mathbf w}
\newcommand{\br}[0]{\mathbf r}
\newcommand{\bR}[0]{\mathbf R}
\newcommand{\bx}[0]{\mathbf x}
\newcommand{\by}[0]{\mathbf y}

\newcommand{\bdelta}[0]{\boldsymbol \delta}
\newcommand{\bbeta}[0]{\boldsymbol \beta}
\newcommand{\bra}[1]{\langle #1|}
\newcommand{\ket}[1]{|#1\rangle}
\newcommand{\braket}[2]{\langle #1|#2\rangle}

\newcommand{\spl}[1]{\begin{align}\begin{split} #1 \end{split} \end{align}}
\newcommand{\al}[1]{\begin{align} #1 \end{align}}
\newcommand{\vvec}[2]{\begin{pmatrix} #1 \\ #2 \end{pmatrix}}
\newcommand{\matr}[4]{\begin{pmatrix} #1 & #2 \\ #3 &  #4 \end{pmatrix}}

\newcommand{\pd}[2]{\frac{\partial #1}{\partial #2}}

\newcommand{\NI}[0]{{N}}

\newcommand{\mI}[0]{m_{\mbox{\tiny I}}}

\newcommand{\omegaI}[0]{\omega_{I}}

\newcommand{\Hph}[0]{H_{\mbox{\tiny ph}}}
\newcommand{\mHph}[0]{\mH_{\mbox{\tiny ph}}}
\newcommand{\mHDelta}[0]{\mH_{\mbox{\tiny $\Delta$}}}
\newcommand{\dl}[0]{l}	

\newcommand{\Dis}{{\rm dist}}

\newcommand{\pic}[3]{
\begin{figure}[tb!]
\includegraphics[width=#3]{#1} \caption  {#2}
\end{figure}
}

\newcommand{\pich}[3]{
\begin{figure}[tbh!]
\includegraphics[width=#3]{#1} \caption  {#2}
\end{figure}
}

\newcommand{\refe}[1]{Eq.~(\ref{EQ:#1})}
\newcommand{\refs}[1]{Sec.~\ref{SEC:#1}}
\newcommand{\reff}[1]{Fig.~\ref{FIG:#1}}

\newcommand{\affA}{Singapore University of Technology and Design, 138682 Singapore}
\newcommand{\affB}{Department of Nuclear Science and Engineering and Research Laboratory of Electronics, Massachusetts Institute of Technology, Cambridge, MA, USA}
\newcommand{\affC}{Institut f\"ur Theoretische Physik, Johann Wolfgang Goethe-Universit\"at, 60438 Frankfurt/Main, Germany}

\begin{document}

\title{Operator-based derivation of phonon modes and characterization of correlations for trapped ions at zero and finite temperature}

\author{U.~Bissbort}\affiliation{\affA}\affiliation{\affB}
\email{ulf.bissbort@gmail.com}
\author{W.~Hofstetter}\affiliation{\affC}
\author{D.~Poletti}\affiliation{\affA}

\date{\today}

\begin{abstract}
We present a self-contained operator-based approach to derive the spectrum of trapped ions. This approach provides the complete normal form of the low energy quadratic Hamiltonian in terms of bosonic phonons, as well as an effective free particle degree of freedom for each spontaneously broken spatial symmetry. We demonstrate how this formalism can directly be used to characterize an ion chain both in the linear and the zigzag regimes. In particular we compute, both for the ground state and finite temperature states, spatial correlations, heat capacity and dynamical susceptibility. Last, for the ground state which has quantum correlations, we analyze the amount of energy reduction compared to an uncorrelated state with minimum energy, thus highlighting how the system can lower its energy by correlations. 
\end{abstract}


\maketitle

\section{Introduction}\label{SEC:Intro}  

Progress in recent decades in quantum optics and trapped ion systems have provided a versatile platform to investigate a broad range of physical phenomena. At sufficiently low temperatures, and depending on the interactions, geometry and parameters of the trap, the contained ions form a variety of  crystalline structures \cite{Raizen1992,Waki1992,Birkl1992,Kjaergaard2003,Piacente2003,Dubin1993,Dubin1997,Horak2012,Cartarius2013,Dessup2015,Dessup2015a}. Particular focus has been cast upon the transition between the linear and the zigzag configurations for a long 1D ion chain, both experimentally \cite{Raizen1992,Waki1992,Birkl1992,Kjaergaard2003} and theoretically \cite{Schiffer1993,Fishman2008,Astrakharchik2008,Gong2010,Baltrusch2011,Shimshoni2011,Shimshoni2011a,Silvi2013,Silvi2014,Podolsky2014}. 

In the context of analog quantum simulation and realization of the idealized models and phenomena previously restricted to textbooks, trapped ion systems have established themselves as one of the major contributors. The realization of spin systems with tunable interactions \cite{Islam2011}, the simulation of relativistic Klein tunneling \cite{Gerritsma2011}, friction at the nanoscale \cite{Benassi2011,Bylinskii2015,Gangloff2015} and the realization of the Jaynes-Cummings model \cite{Leibfried2003} are only a few prominent examples. Also in the context of time-dependent, out of equilibrium physics, these systems have been used to study the transport of energy and particles \cite{Bermudez2013,Ruiz2014,Ramm2014,Guo2015,Guo2016}, the scaling of defects formation by ramping across the zigzag transition \cite{Ulm2013,Baltrusch2013} and provided an experimental test of the Jarzynski equality in the quantum regime \cite{An2014}.

Recently, proposals for the realization of systems including both ions and neutral atoms have been investigated \cite{Bissbort2013,Secker2016}, leading the way to new avenues for research of complex quantum systems. 

In all of the above systems the phonons modes between different ions play an essential role in realizing the properties of the respective system, be it by providing a coupling between the relevant degrees of freedom for spins or being the object of study themselves. Also in the context of digital quantum computation with trapped ions \cite{Lanyon2011,Kielpinski2002}, where the qubits are typically encoded in the ions' hyperfine states, the phonon modes provide the bus via which information can be exchanged and are essential in the realization of single- and two-qubit gates. 
A detailed understanding of the phononic structure in ion systems is therefore important for the research directions discussed above \cite{James1998,Home2011,Home2011,Baltrusch2012,Feldker2014}. In this work we provide an alternative method to calculate the phonon structure compared to established ones \cite{James1998, Home2011, Baltrusch2012, Dessup2015}, where one determines the collective eigenmodes of the classical system and subsequently quantizes these. Instead, starting from the microscopic exact Hamiltonian and remaining on the full operator level throughout, we completely describe the system from first principles with both bosonic phononic and non-bosonic zero energy degrees of freedom. Our approach features strong analogy to Bogoliubov and BCS theory, allowing a number of well-known techniques to be directly adapted. We demonstrate how various quantities such as real space correlations, heat capacity and susceptibility can be readily computed both for the ground state and at finite temperature, intrinsically accounting for the correlations contained in the many-body eigenstates. 

\pic{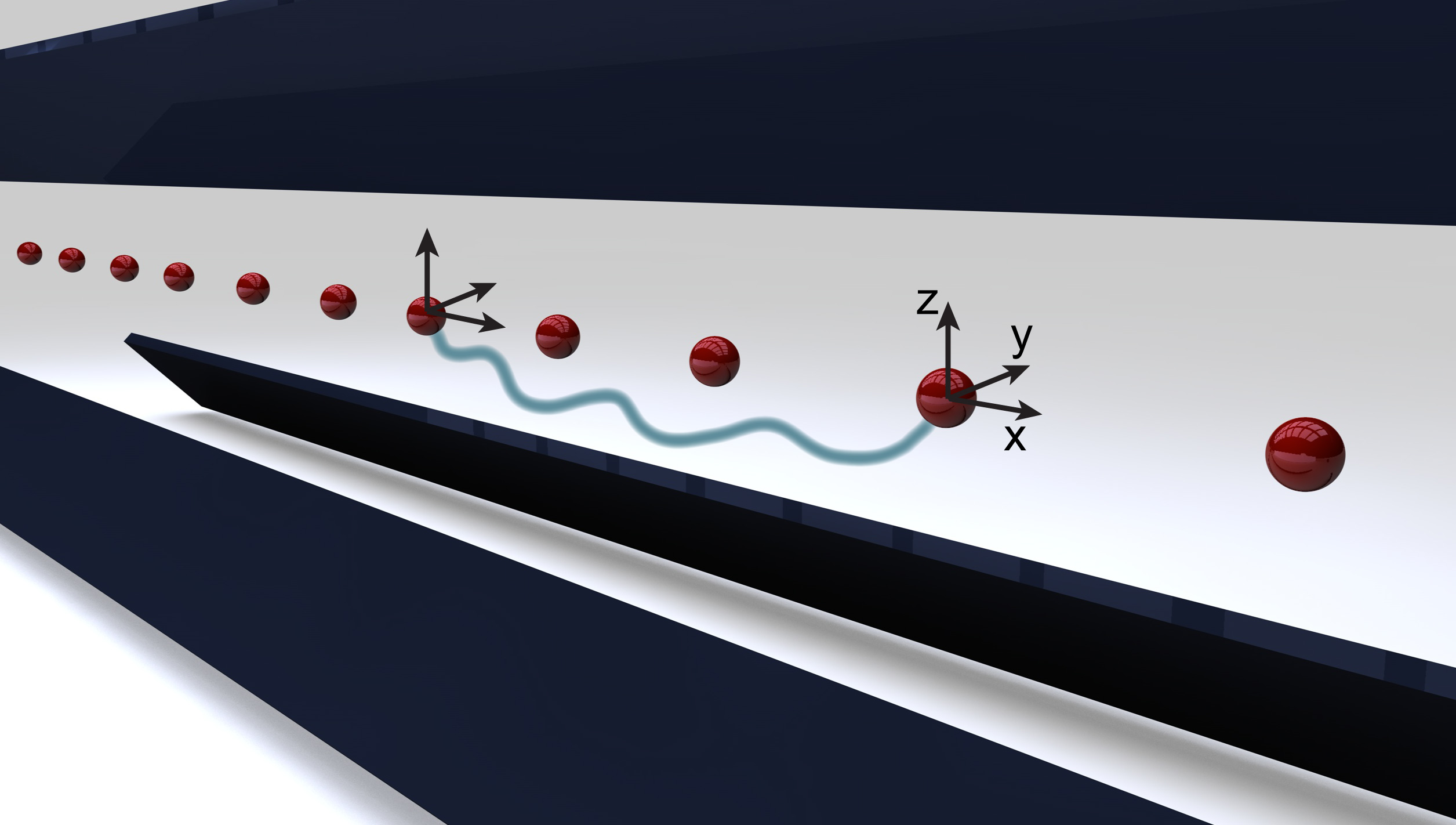}{In a 1D trapped ion system (shown here in the linear chain configuration in a Paul trap), the motion and position of the individual ions are generally entangled. The correlations are intrinsically contained in the phonon ground state $\ket{\psi_0}$ and are explicitly calculated in \refs{spatial_correlations}.}{\linewidth}

In \refs{PhononTransformation} we present in detail a general procedure to write the Hamiltonian in diagonal form. We then apply this formalism to an ion chain in  \refs{applionchain}, concentrating on the linear case in  \refs{linear_chain} and to the zigzag regime in \refs{zigzag_chain}. The structure of the effective free particle degrees of freedom is discussed in \refs{zeromode}, as well as the regime of validity using the Ginzburg criterion in \refs{Ginzburg_crit}. We then apply our formalism to compute real-space correlations, heat capacity and susceptibility respectively in Sections~\ref{SEC:spatial_correlations}, \ref{SEC:heat_capacity} and \ref{SEC:local_susceptibility}, as well as the correlation energy reduction in \refs{correlation_energy_reduc}. Finally we draw our conclusions in \refs{conclusions}.

\section{General Phonon Transformation on the Operator Level}\label{SEC:PhononTransformation}

Let us consider a system of $N$ ions in a one-dimensional linear configuration (although the formalism applies to systems of arbitrary dimension), specified by their position and momentum operators $\{\bR_l\}$ and $\{\bP_l\}$ and assume that the Hamiltonian is of the form \footnote{This assumption is not essential for the method and it can be extended in a straight-forward way to systems with cross-coupling terms between $\bR$ and $\bP$.}
\spl{
\label{EQ:H_form_general}
\mH&=\sum_{l,\md} \frac{1}{2 m_{l}} {\bP_{l}}^2 + V(\{ \bR_{l} \}),
}
where $m_l$ is the mass of the $l$-th ion. We decompose each position operator into a scalar equilibrium position $\overline \bR_l$ and a displacement operator $\bR_l=\overline \bR_l + \delta \bR_l$. Working in units $\hbar=1$, these position fluctuation operators fulfill the same elementary commutation relations $[\delta \bR_{l,\md}, \, \bP_{l',\md'}] = i \delta_{l,l'} \delta_{\md,\md'}$, where $\md,\md' \in \{x,y,z\}$ are dimensional indexes and $i$ is the imaginary unit. The equilibrium position can be chosen as the classical equilibrium position or self-consistently, as a variational parameter, for instance within a Hartree-Fock Bogoliubov approach. For the infinite linear chain or a system with periodic boundary conditions, the symmetry dictates an equidistant spacing \footnote{In case the reader feels uncomfortable with parametrically fixing $d$ in an infinite chain (somewhat analogous to fixing the total particle number in condensed matter models), one may also consider the ion in a potential $V(z)= \tilde \mu|z|$ and perform the calculation at fixed $\tilde \mu$. In the bulk, this leads to a constant $d(\tilde \mu)$. Being a strictly decreasing function, one finds a one to one correspondence between the two quantities and can adjust $\tilde \mu$ for a required $d$ accordingly.} $\overline \bR_l=l d \be_x$, where we parametrically fix the inter-ion spacing $d$.

To determine the phonon operators, phonon frequencies and additional effective position and momentum operators to complete the operator basis, we expand the Hamiltonian  in the operators $\delta \bR_{l,\md}$ and $\bP_{l',\md'}$ around a parametrized equilibrium $\overline \bR_l$, $\overline \bP_l=0$. The potential is to be understood as an operator throughout and the Taylor expansion coefficients are identical to those in the expansion of the classical potential energy function.

Note that the relevance of the quantum statistical properties of the ions, i.e. whether the (anti-) symmetry of the many-body ion state leads to any modification in observables, is governed by the ratio of the longitudinal harmonic oscillator length and the inter-ion spacing $d$. For any experimentally realistic set of conditions this ratio is very small \cite{Eschner2003}, justifying the treatment of the different ions as distinguishable particles to a very high degree.

Formally, one can write the Hamiltonian as a sum of terms of different order
\al{
\mH&=\mHph+\mHDelta\\
\mHph&=E_0+\mH^{(1)}+\mH^{(2)} \label{EQ:H_ph_def}\\
\mHDelta&=\sum_{m=3}^\infty  \mH^{(m)}
}
where $\mH^{(m)}$ contains products of exactly $m$  factors of operators $\delta \bR_{l,\md}$ and/or $\bP_{l',\md'}$ and $\mHph$ contains all terms up to second order. Expanding around the classical equilibrium, $E_0$ is simply the classical electrostatic energy and a constant off-set. $\mH^{(1)}$ vanishes, as otherwise a change of the positions/momenta of some ions could reduce the energy in linear order in the displacement. The $\mH^{(m)}$ terms are not invariant under a shift of the scalar offset around which the expansion is performed, i.e. mixing between the terms of different orders occurs if the expansion is done around a different minimum.

All terms up to second order can be brought into diagonal form by a generalized unitary (Bogoliubov-type) transformation. The higher order terms $\mH_\Delta=\sum_{m=3}^\infty \mH^{(m)}$ subsequently describe possible phonon interactions and decay processes, discussed for instance in \cite{Home2011,Marquet2003,Feldker2014}.

To determine the phonon structure, we focus on $\mHph$, the phonon part Eq.~(\ref{EQ:H_ph_def}) which is at most quadratic in the momentum and position operators. Instead of working directly in terms of the $\delta \bR_{l,\md}$ and $\bP_{l,\md}$ operators, we first introduce the local harmonic oscillator ladder operators $b_{l,\md}$ and $b_{l,\md}^\dag$. 

The phonon transformation can be defined as the generalized unitary basis transformation acting on the space of the bilinear coefficients of the ladder operators $b_{l,\md}$ and $b_{l,\md}^\dag$, that brings the $\mHph$ into diagonal form, while preserving the bosonic commutation relations \footnote{Mathematically, such a phonon transformation is equivalent to a multi-mode squeezing (together with unitary rotations and shifts) transformation and also analogous to Lorentz transformations, which preserve a similar symplectic pseudo-norm and the generators can subsequently be classified into rotations and boosts.}. We begin by defining the bare, local harmonic oscillator frequencies
\al{
\label{EQ:bare_ho_freqs}
\Omega_{l,\md}&=\sqrt{\frac{V_{(l,\md),(l,\md)}  }{m_l}},
} 
obtained by expanding the potential $V(\{\bR_l\})$ around the equilibrium position $\overline \bR$ 
\spl{
\label{EQ:potential_expansion_coeff}
V_{(l,\md),(l',\md)} := \left. \frac{\partial^2 V}{\partial \bR_{l,\md}\, \partial \bR_{l',\md'}}\right|_{\overline \bR}.
}
In a system with discrete translational symmetry, $\Omega_{l,\md}$ does not depend on the ion index $l$ and we omit this henceforth. We now define ladder operators for the local harmonic oscillators
\al{
b_{l, \md}&=\sqrt{\frac{ m_l \Omega_{(l,\md)}} {2} }\Big[\delta \bR_{l,\md} + \frac{i}{m_l \Omega_{l,\md}} \bP_{l,\md} \Big],
}
with the inverse transformation 
\al{
\label{EQ:trafo_deltaX_in_b}
\delta \bR_{l,\md}&=\sqrt{\frac{1}{2 m_l \Omega_{l,\md}}}(b_{l, \md}+b_{l, \md}^\dag)\\
\bP_{l,\md} &=i\sqrt{\frac{ m_l \Omega_{l,\md}}{2}}(b_{l, \md}^\dag-b_{l, \md} )
}
allowing the position fluctuation and momentum operators of the ions to be expressed in terms of ladder operators $b_{l, \md}$ and $b_{l, \md}^\dag$. Subsequently, $\mHph$ can be written as 
\spl{
\label{EQ:H_I_second_order_in_b}
\mHph &\approx E_0+ \frac 1 2 \sum_{l, \md} \Omega_{l,\md} [b_{l,\md}^\dag b_{l,\md}+ b_{l,\md} b_{l,\md}^\dag] \\
&+ \frac 1 2 \sum_{\stackrel{(l,\md),(l',\md')}{{\tiny(l,\md)\neq (l',\md')}}} \frac{V_{{(l,\md)},{(l',\md')}}}{2m_l \sqrt{ \Omega_{l,\md} \Omega_{l',\md'} } } \Big[
b_{l,\md}^\dag b_{l',\md'} + b_{l,\md}  b_{l',\md'}^\dag \\
& + b_{l,\md} b_{l',\md'} + b_{l,\md}^\dag b_{l',\md'}^\dag \Big],
}
which is a bilinear form in the creation and annihilation operators.  Defining the column vector operators 
\spl{
\label{EQ:b_column_vecs}
\bb=\begin{pmatrix}  b_{1,x} \\ b_{2,x}  \\ \vdots \\ b_{N,z} \end{pmatrix}, \qquad 
\bb^\dag=\begin{pmatrix}   b_{1,x}^\dag \\ b_{2,x}^\dag  \\ \vdots \\ b_{N,z}^\dag \end{pmatrix},
}
$\mHph$ can be expressed in compact notation as
\spl{
\label{EQ:H_I_quadratic_in_bs}
\mHph&\approx E_0+ \frac 1 2 {\vvec{\bb}{\bb^\dag}}^\dag    \Hph  \vvec{\bb}{\bb^\dag}.
}
Here the phonon quasi-particle Hamiltonian matrix of coupling elements $\Hph$ is 
\spl{
\label{EQ:H_phonon}
\Hph=\matr{h}{g }{g^*}{h^*}
}
of which the matrix elements are
\spl{
{g}_{l,m}^{(\dl)}&=(1-\delta_{l,m})\frac{V_{l,\md}}{2 m_l \Omega_{(\dl)} }\\
}
\spl{
\label{EQ:h_def}
h_{l,m}^{(\dl)}&=  \delta_{l,m} \, \Omega_{(\dl)} + (1-\delta_{l,m})\frac{V_{l,m}}{2 m_l \Omega_{(\dl)} } \\
&=\delta_{l,m} \, \Omega_{(\dl)} + g_{l,m}^{(\dl)}.
}
By virtue of Schwarz's theorem, the smoothness of the Coulomb potential at equilibrium and the potential $V(\{\bR_l\})$ being real, the matrices $\mathbbm{V}$ (with elements $V_{l,m}$), $g$ and $h$ are symmetric.

\subsection{Diagonalizing transformation}
\label{SEC:Transformation}
Given the quasi-particle bilinear coupling matrix $\Hph$ (not to be confused with the many-body operator $\mHph$), we shall now discuss how to generally obtain the phonon transformation leading to a representation in terms of non-interacting quasi-particles and effective free particle degrees of freedom in the presence of gapless modes. It is useful to map this problem onto a diagonalization and subsequent completion problem instead of imposing the bosonic commutation relations. The latter approach does not generally lead to a complete operator basis, as we will show later and becomes tedious for systems lacking a high degree of symmetry, such as spatially inhomogeneous systems. 

As $\Hph$ is Hermitian, it may seem tempting to directly diagonalize this matrix, 
however the orthogonality between the eigenvectors with respect to the standard Euclidean scalar product is at variance with the canonical structure of the transformations and would not preserve the commutation relations.
 Rather, the diagonalization has to be performed on a symplectic space, where a (pseudo-) scalar product $(\bx | \by):= \bx^\dag \Sigma \by$ and an induced (pseudo-) norm\footnote{the square is to be understood symbolically} $||\bx||^2 := (\bx| \bx)$ are defined by the matrix 
\spl{
\Sigma=\matr{\mathbbm{1}_\mD}{0}{0}{-\mathbbm{1}_\mD}.
}
Here $\mathbbm{1}_\mD$ is the identity matrix on a space of dimension $\mD$ given by the number of bare bosonic degrees of freedom \footnote{The same scalar product and properties appear in the Hamiltonian formulation of classical mechanics, albeit typically expressed in a different basis.}. In later sections, $\Sigma$ denotes the same matrix in reduced dimensions, the latter being defined implicitly with the diagonal submatrices $\mathbbm{1}_\mD$ and $-\mathbbm{1}_\mD$ being of equal dimension, as usual on symplectic spaces.

First, we note that $\Sigma \Hph$ is no longer Hermitian (nor normal, i.e. $[\Sigma \Hph, \, (\Sigma \Hph)^\dag ]\neq 0$ and a basis of eigenvectors does not generally exist. 
In general, any $2\mD$-dimensional vector can be classified into three qualitatively distinct classes based on its $\Sigma$-norm: positive, negative or zero, a property which is invariant under any renormalization or basis representation. In fact it follows from the general structure of $\Sigma \Hph$ that positive and negative vectors (in the norm) always appear in pairs: for every positive (right) eigenvector $\bx^{(m)}=\vvec{\bu^{(m)}}{-\bv^{(m)}}$ with an eigenvalue $\omega_m$, a conjugate counterpart 
\spl{
\by^{(m)}=\vvec{-\bv^{(m)^*}}{\bu^{(m)^*}}=\matr{0}{\mathbbm 1}{\mathbbm 1}{0} \bx^{(m)^*}
}
with a negative norm also appears as a (right) eigenvector of $\Sigma \Hph$ to an eigenvalue $-\omega_m^*$. We have chosen to label every eigenvector and eigenvalue by a combined index $m$, which in the case of symmetries may be decomposed into groups of quantum numbers [such as the quasi-momentum and direction in the linear chain $m=(k,\md)\,$]. Each positive eigenvector $\bx^{(s)}$ is associated with an annihilation operator 	
\spl{
\label{EQ:def_beta}
\beta_m= {\bx^{(m)}}^\dag \Sigma \vvec{\bb}{\bb^\dag},
} 
and the conjugate eigenvector with a negative norm $\by^{(m)}$ is associated with the corresponding creation operator 
\spl{
\label{EQ:def_betadag}
\beta_m^\dag= -{\by^{(m)}}^\dag \Sigma \vvec{\bb}{\bb^\dag}.}  
Care has to be taken in choosing the global complex phases of the eigenvectors $\bx^{(m)}$ and $\by^{(m)}$ suitably for these properties to be consistent. 
One can now relate the properties of the operators $\beta_m^\dag$ and $\beta_m$ to purely algebraic properties of the associated vectors. Specifically, for arbitrary  vectors $\bv, \, \bw \in \mathbb C^{2\mD}$, the associated operators $\hat V := \bv^\dag \Sigma \vvec{\bb}{\bb^\dag}$ and $\hat W := \bw^\dag \Sigma \vvec{\bb}{\bb^\dag}$ fulfill the commutation relations  
\spl{
[\hat V, \hat W^\dag]=(\bv | \bw)
}
where the right side is implicitly understood as multiples of the unit operator on the operator space. Thus the orthogonality $(\bx_m | \bx_{m'})=\delta_{m,m'}$ corresponds to the orthogonality of the modes on a single-particle level and choosing the normalization $(\bx_m| \bx_{m})=1$, $(\by_m| \by_{m})=-1$ leads to the correct bosonic normalization of the commutator $[\beta_m, \beta_m^\dag]=1$ and $[\beta_m^\dag, \beta_m]=-1$ respectively. Generally, the eigenvectors to different eigenvalues are inherently $\Sigma$-orthogonal, thus a normalization and phase matching after diagonalization is sufficient. In case of degeneracies an additional $\Sigma$-orthogonalization has to be performed.

Usually one finds eigenvectors with a positive $\Sigma$-norm associated with positive eigenvalues and vice versa. This corresponds to thermodynamic stability of the system (for a discussion, see e.g. \cite{Castin2001}), but there is no fundamental guarantee for this property. If this is not the case, the system can lower its energy by populating phonon modes. Note that thermodynamic instabilities are different from dynamical instabilities, where the eigenvalues $\omega_m$ become complex. 

\subsection{The zero-subspace}
\label{SEC:properties_zero_subspace}
In the case of non-zero, non-degenerate and real eigenvalues, the existence of linearly independent eigenvectors is guaranteed (as the eigenvalues appearing in  pairs are different and each one possesses an associated eigenvector). 
The eigenvector to a zero eigenvalue $\Sigma \Hph \bp=0$ can generally be chosen to be of the form 
\spl{
\label{EQ:p_phase_form}
\bp=\vvec{\bu^{(0)}}{-{\bu^{(0)}}^*},
}
which guarantees the associated operator 
\spl{
\label{EQ:def_P}
\mathscr P = {\bp}^\dag \Sigma \vvec{\bb}{\bb^\dag}
} 
to be Hermitian. A second, linearly independent eigenvector to zero does not exist and one needs an additional vector $\bq$ to complete the basis. From a mathematical perspective the most general normal form one can bring the matrix into is the Jordan normal form and $\bq$ is a generalized eigenvector to the sub-block with eigenvalue zero. Hence $\bq$ is not a true eigenvector, and is mapped onto $\bp$ by    
\spl{
\label{EQ:mapping_q_onto_p}
\Sigma \Hph \, \bq = - \frac{i}{\tilde m}  \bp,
}
where an effective (purely real) mass $\tilde m=1/||\Sigma \Hph \, \bq||_2$ emerges, which is not to be confused with the effective mass of emerging quasi-particles (the inverse respective band curvature). The choice of using `$i$' in this property is consistent with fixing the the scalar product
\spl{
\label{EQ:QP_p_q_overlap}
(\bq | \bp)=\bq^\dag \Sigma \bp=i,
}
which reflects the associated effective position operator 
\spl{
\label{EQ:Q_def}
\mathcal Q = -{\bq}^\dag \Sigma \vvec{\bb}{\bb^\dag}.
}
being conjugate to $\mathscr P$, i.e. $[\mathcal Q,\mathscr P]=i$. A closer inspection shows that $\bq$ can be chosen to be of the form 
\spl{
\label{EQ:q_phase_form}
\bq=-i\vvec{\bv^{(0)}}{{\bv^{(0)}}^*},
}
where it turns out that the elements of the subvectors $\bv^{(0)}$ and $\bu^{(0)}$ are purely imaginary.
Note that Eq.~(\ref{EQ:mapping_q_onto_p}) and \refe{QP_p_q_overlap} do not uniquely determine the normalization $\bq$ and $\bp$. Since $\bp^\dag \Sigma \bp =\bq^\dag \Sigma \bq  = 0$, neither of these vectors can be $\Sigma$-normalized and we have to resort to another normalization to uniquely characterize them. We first point put that the choice of normalization of $\bp$ and $\bq$ is arbitrary as long as $||\bp||_2=1/||\bq||_2$ and this choice is equivalent to choosing a length scale for the center of mass degree of freedom, fixing the inverse scale for the conjugate momentum. To retain the analogy with the conventional choice of center of mass coordinates and total momentum, we choose the normalization $\bp^\dag \bp=\NI$ and $\bq^\dag \bq=1/\NI$. This corresponds to $\mQ$ being a mean position (i.e. an intensive quantity) and $\mathscr P$ being the total momentum (i.e. an extensive quantity). Furthermore, with this choice, $\tilde m$ is extensive. This is to be expected on a simple classical level, where the total mass of the particles appears in the center of mass contribution to the kinetic energy in conjunction with the total physical momentum.

Together, these conditions uniquely determine $\bp$, $\bq$ and $\tilde m$ up to an overall sign in both vectors. For both the longitudinal and helical effective free particle degree of freedom, the elements of the vectors $\bv^{(0)}$ and $\bu^{(0)}$ turn out to be purely imaginary. Note that Eq.~(\ref{EQ:QP_p_q_overlap}) directly implies the commutation relations $[\mathcal Q, \mathscr P]=i$ in accordance with the analogy to an effective single particle. This natural expression is a result of the choice of using `$i$' in Eq.(\ref{EQ:mapping_q_onto_p}). The commutators automatically fulfill $[\beta_m, \, \mathcal Q]=0$, since the associated eigenvectors are in different eigenspaces and thus mutually $\Sigma$-orthogonal. We also point out that in contrast to the phonon vectors $\bx$ and $\by$, the form of Eqs.~(\ref{EQ:p_phase_form},\ref{EQ:q_phase_form}) fixes the global complex phase of the vectors $\bp, \bq$ up to a minus sign. The fact that the complex phase of these vectors is not arbitrary will be of significance later, where observables may depend on it, as for instance in Eqs.~(\ref{EQ:akd_ak},\ref{EQ:akd_akd}) when calculating spatial correlations.

\subsection{Completeness and normal form}
Since the eigenvectors of $\Sigma \Hph$ do generally not form a complete basis, its normal form is not given by the usual expansion in eigenvectors only. A more general, systematic procedure to obtain the full Jordan normal form is required, which consists of constructing the completeness relation on the $2\mD$-dimensional space in terms of the vectors $\{\bx, \, \by, \, \bp, \, \bq \}$ \cite{Blaizot86,Lewenstein1996,Castin2001}. Subsequently this is inserted into the expression for $\mH^{(2)}$, guaranteeing that all degrees of freedom are accounted for. Once this is established, we formulate the basis transformation between the original ladder operators and the resulting phonon, $\mathscr P$ and $\mathcal Q$ operators in a compact way by defining generalized unitary matrices and exploiting their properties. As not to clutter the notation, we will formulate the construction for the case of a single gapless mode, but the extension to multiple (or no) zero modes is a natural extension by using multiple (or no) $\bp$ and $\bq$ vectors. 

We construct a completeness relation, i.e. construct the $2\mD$-dimensional unit matrix, in terms of vectors $\{\bx, \by,\bp,\bq\}$. since any vector can thus be uniquely decomposed and expressed in terms of these vectors (the coefficients can be directly be determined using the $\Sigma$-orthogonality), this relation implies that any bare creation and annihilation operator $b$, $b^\dag$ (or equivalently any ion's position or momentum operator) can be expressed in terms of the operators $\beta_m, \, \beta_m^{\dag}, \mathcal Q, \mathscr P$. Since any many-body operator can be expressed in terms of a sum of products of $b$ and $b^\dag$ (i.e. these are the generators of the group of many-body operators), $\{\beta_m, \, \beta_m^{\dag}, \mathcal Q, \mathscr P \}$ constitutes an equally valid set of generators in terms of which any many-body operator can be expressed.

Before writing the completeness relation, it is useful to summarize a number of relations that these vectors fulfill, which are 
  \spl{
	\label{EQ:orthog}
	({\bx^{(r)}} |  \bx^{(s)} )&= \delta_{r,s}\\
	({\by^{(r)}} | \by^{(s)})&= -\delta_{r,s}\\
	({\bx^{(r)}} | \by^{(s)})&=0\\
	(\bp \,| \bp )&= 0\\
	(\bp \,| \bx^{(s)}) &= 0\\
	(\bp \,| \by^{(s)}) &= 0\\
	(\bq \,| \bq )&= 0\\
	(\bq \,| \bx^{(s)} )&= 0\\
	(\bq \,| \by^{(s)} )&= 0\\
	(\bq \,| \bp) &= i\\
	\bp^\dag  \bp &= 1.
	}

We postulate that the completeness relation in terms of these vectors now reads (see e.g. \cite{Blaizot86})
\spl{
\label{EQ:completeness_total}
\mathbbm{1}_{2\mD}&= \sum_{m} (\bx^{(m)} {\bx^{(m)}}^\dag - \by^{(m)} {\by^{(m)}}^\dag)\Sigma + \; i (\bq \, \bp^\dag - \bp \, \bq^\dag)\Sigma.
}
This relation can be proven by multiplying the right side of Eq.~(\ref{EQ:completeness_total}) by any of the basis vectors and, using the orthogonality relations in Eq.~(\ref{EQ:orthog}), establishing that each of these vectors is mapped onto itself. Asserting that all the vectors are linearly independent and their number equals the dimension of the space, it follows that they span a $2\mD$-dimensional space, establishing the validity of Eq.(\ref{EQ:completeness_total}) as a completeness relation. For multiple effective free particle degrees of freedom, the last term again has to be extended accordingly.

The representation of $\mH^{(2)}$ given in Eq.~(\ref{EQ:H_I_quadratic_in_bs}) can now be brought into normal form by inserting the identity in form $\mathbbm 1_{2 \mD} = \Sigma^2$ twice: once on the left of $\Hph$ and once on the right in form of Eq.~(\ref{EQ:completeness_total}). Restricting to dynamically stable systems and using the eigenvector relations  $\Sigma \Hph \bx^{(s)} = \omega_s  \bx^{(s)}$, $\Sigma \Hph \by^{(s)} = -\omega_s  \by^{(s)}$ and $\Sigma \Hph \bq=0$ one finds
\spl{
\label{EQ:H2_diagonal}
\mH^{(2)} =&  \frac 1 2 {\vvec{\bb}{\bb^\dag}}^\dag \Sigma \big[\sum_m  ( \bx^{(m)} \omega_m \bx^{(m)^\dag}  - \by^{(m)} (-\omega_m) \by^{(m)^\dag}   ) 
 \\  &+\frac{1}{\tilde m} \bp \bp^\dag \big] \Sigma     \vvec{\bb}{\bb^\dag}\\
=& \sum_m  \omega_m \beta_m^\dag \beta_m + \frac{\mathscr P^2}{2 \tilde m} + \Delta E^{(2)}.
}
In getting to the last line, we have used the definition of the phonon operators in Eq.~(\ref{EQ:def_beta}) and the effective momentum operator in Eq.~(\ref{EQ:def_P}), which justifies the previous definition. From this form the analogy of $\mathscr P$ and $\mathcal Q$ describing an effective free particle also becomes clear, where $\tilde m$ plays the role of a mass. Note that $\mH^{(2)}$ does not fix the boundary condition(s) for the effective free particle degree(s) of freedom and this must be specified additionally. 

Furthermore, a zero point energy shift 
\spl{
\label{EQ:zero_pt_en_shift}
\Delta E^{(2)}=\frac 1 2 \sum_m \omega_m 
}
appears in the last line of Eq.~(\ref{EQ:H2_diagonal}), which originates from the reordering of bosonic operators when bringing $\mH^{(2)}$ into the final form. 

\subsection{Transformation relations between local modes and normal modes of the Hamiltonian}

It is now important to detail the relation between the local modes $\bb$ and $\bb^\dag$ with the normal modes of (\ref{EQ:H2_diagonal}), i.e. to obtain compact transformation relations between the vectors of operators
\spl{
\label{EQ:phonon_op_trafo}
\begin{pmatrix} 
\bbeta\\
\mathcal Q\\
\bbeta^\dag\\
\mathscr{P}
\end{pmatrix} \longleftrightarrow
 \vvec{\bb}{\bb^\dag}. 
} 

We begin by defining the matrix $W$ by arranging the eigenvectors and $\bq$ as column vectors
\spl{
\label{EQ:Bog_W_matrix_def}
W=\Big[  \bx^{(1)}, \ldots ,\bx^{(\mD-1)},\,  i  \bp,\, \by^{(1)}, \ldots ,\by^{(\mD-1)},  i  \bq  \Big].
}
The factors $i$ and the ordering are chosen such that the resulting transformations take on a particularly simple form. Since all column vectors are linearly independent, the matrix $W$ is non-singular and the inverse $W^{-1}$ exists. Next, we define $\tilde\Sigma$ as a useful matrix which is given by the product      
\spl{
\label{EQ:Sigma_tilde_def_Bog}
\tilde \Sigma := W^\dag \Sigma W = 
\begin{pmatrix} 
\mathbbm{1}_{\mD-1} \\
 & 0  & \ldots &  -i\\
 &\vdots & -\mathbbm{1}_{\mD-1} &  \vdots \\
 & i  & \ldots &  0
\end{pmatrix},
}
where we used the relations in Eq.~(\ref{EQ:orthog}) to determine the individual matrix elements. $\tilde\Sigma$ is a Hermitian and unitary matrix, in fact $\tilde \Sigma=\tilde \Sigma^\dag$ fulfilling $\tilde \Sigma^2=\mathbbm{1}$. In presence of multiple gapless modes, $\tilde \Sigma$ can be modified by replacing the $i$'s by multiples of the unit matrix in the zero subspace.
By multiplying both sides of Eq.~(\ref{EQ:Sigma_tilde_def_Bog}) by $\tilde \Sigma$ from the left, one finds $\tilde \Sigma W^\dag \Sigma W  = \mathbbm{1}$.
This directly gives us access to the inverse matrix $W^{-1}$ without explicit inversion (the left and right inverse are identical on any finite dimensional space)
\spl{
\label{EQ:Bog_inv_W}
W^{-1}=\tilde \Sigma W^\dag \Sigma=
\begin{pmatrix} 
U^\dag, & V^\dag \\
-i {\bv^{(0)}}^\dag, & i {\bv^{(0)}}^t\\
V^t, & U^t\\
{\bu^{(0)}}^\dag, &  {\bu^{(0)}}^t\\
\end{pmatrix}
}
and $V^t$ denotes the matrix transpose of $V$. To determine the transformation, we use
\spl{
\label{EQ:Bog_op_trafo_intermediate}
\begin{pmatrix} 
\bbeta\\
-i \mathscr{P}\\
-\bbeta^\dag\\
i \mathcal Q
\end{pmatrix} =W^\dag \Sigma 
 \vvec{\bb}{\bb^\dag},
}
which can easily be verified by explicit multiplication and comparison with the respective operator definitions. Multiplying Eq.~(\ref{EQ:Bog_op_trafo_intermediate}) by $\tilde \Sigma$, one finds the explicit form of both the forward and backward transformation of operators
\al{
\label{EQ:Bog_op_trafos_matrix}
\begin{pmatrix} 
\bbeta\\
 \mathcal Q\\
\bbeta^\dag\\
\mathscr{P}
\end{pmatrix} 
&= \tilde \Sigma W^\dag \Sigma  \vvec{\bb}{\bb^\dag} = W^{-1} \vvec{\bb}{\bb^\dag}\\
\label{EQ:Bog_op_trafos_matrix_reverse}
 \vvec{\bb}{\bb^\dag}&=W
\begin{pmatrix} 
\bbeta\\
 \mathcal Q\\
\bbeta^\dag\\
\mathscr{P}
\end{pmatrix}.
}
Since any many-body operator can be expressed in terms of creation and annihilation operators (possibly containing multiple products of these), it can also be expressed exactly in terms of operators $\{\bbeta, \, \bbeta^\dag, \, \mathscr{P}, \, \mathcal{Q}  \}$. 

For completeness, we give the explicit form of the reverse transformation in Eq.~(\ref{EQ:Bog_op_trafos_matrix_reverse}), which reads
\al{
 b_{(l,\md)} &= \sum_m u_{(l,\md)}^{(m)} \beta_m \nonumber\\ 
 &- \sum_m {v_{(l,\md)}^{(m)}}^* \beta_m^\dag \nonumber\\ 
 &+ i u_{(l,\md)}^{(0)} \, \mathcal Q + v_{(l,\md)}^{(0)} \, \mathscr P\\
 b_{(l,\md)}^\dag &= \sum_m {u_{(l,\md)}^{(m)}}^* \beta_m^\dag \nonumber\\
 &- \sum_m {v_{(l,\md)}^{(m)}} \beta_m  \nonumber\\
 &- i {u_{(l,\md)}^{(0)}}^* \, \mathcal Q + {v_{(l,\md)}^{(0)}}^* \, \mathscr P.
}
Here $u_{(l,\md)}^{(m)}$ is to be understood as the $(l,\md)$-th element of the $m$-th eigenvector's subvector $\bu$. We remark that the presence of the $\mathcal Q$ and $\mathscr P$ terms is commonly overlooked when expanding the local ion operators in terms of phonon ladder operators. The former two terms are however essential in constructing a complete operator algebra, accounting for all degrees of freedom.

\section{Application to an Ion Chain} \label{SEC:applionchain}
We will now apply this procedure to a 1D ion chain, both in the linear and the zigzag regimes, as has been considered in \cite{Fishman2008}. 
We consider ions of equal mass $m_l=\mI$, mutually interacting via the Coulomb force and each of them trapped in a radially symmetric trapping potential. The formalism can be readily extended beyond these assumptions. The Paul trap uses a time dependent potential to trap the ions. In the secular approximation it can be described by an effective stationary quadratic potential. The resulting Hamiltonian is thus
\spl{
\label{EQ:ion_Hamiltonian}
\mH_I&=\sum_{l}\frac{\bP_l^2}{2 \mI} + \hat V\\
\hat V &=  \sum_l \frac{\mI \omegaI^2}{2} (\alpha \;  \delta \bR_{l,z}^2 +  \delta \bR_{l,y}^2)+\frac{q^2}{4 \pi \epsilon_0 }\sum_{\stackrel{l,l'}{\mbox{\tiny $l\neq l'$}}} \frac{1}{| \bR_l -  \bR_{l'}|},
}

where $q$ is the charge of each ion and $\omegaI=\omega_y$ is the trapping frequency in the $y$ direction. The trapping frequency in the $z$ direction is $\omega_z=\alpha\omegaI$, where $\alpha$ is the anisotropy of the radial potential ($\alpha=1$ for the radially symmetric case). 

To accommodate for the zigzag transition, we parametrize the position of the $j-$th ions as
\spl{
\label{EQ:equilib_poisitons}
\overline \bR_l=l d \,\be_x + \Delta \; (-1)^l  \be_y,
}
where $\Delta$ is the classical equilibrium displacement. Already on a classical level the system features a transition between the linear ($\Delta = 0$) and zigzag ($|\Delta| > 0$) regimes, determined by the value of $\Delta$ that minimizes the free energy. It is useful to define the dimensionless parameter 
\spl{
\kappa:=\frac{q^2}{2 \pi \epsilon_0 m_I \omega_I^2 d^3},
}
which governs the zigzag transition, as will become apparent later. It may be noted, that $\kappa$ is the ratio of the electrostatic energy $\frac{q^2}{4 \pi \epsilon_0 d}$ of two charges separated by a distance $d$ and the potential energy $\frac 1 2 m_I \omega_I^2 d^2$ of a harmonic oscillator with frequency $\omega_I$ excursed a distance $d$. It is thus the ratio of two competing energy scales along the radial and longitudinal directions. At a critical value $\kappa_c$ a phase transition between a linear ($\kappa<\kappa_c$) to a zigzag ($\kappa>\kappa_c$) chain occurs. 

Defining the natural energy unit for the system $E_d:=\frac 1 2 \mI \omegaI^2 d^2$ and expressing all distances in units of $d$, $\kappa$ becomes the prefactor of the repulsive Coulomb potential, i.e.
\spl{\label{EQ:potentialdelta}
\frac{V}{E_d}=  \sum_l  \left( \frac{\delta \bR_{l,y}^2}{d^2} +  \alpha \; \frac{\delta \bR_{l,z}^2}{d^2} \right)+ \kappa \sum_{\stackrel{l,l'}{\mbox{\tiny $l\neq l'$}}} \frac{1}{| \frac{\bR_l}{d} -  \frac{\bR_{l'}}{d}|} .
}
If all distances are expressed in units of $d$ and the interactions are quantified by the dimensionless coupling strength $\kappa$, $E_d$ sets the natural energy scale of the system.

The classical equilibrium position of the ions is determined by minimizing the classical potential energy (here: per particle), which is given by
\spl{
\frac{V(\Delta)}{\NI\; E_d}=
&\left\{ \left( \frac \Delta d \right)^2 \right.\\    
+ & \left.     \kappa \sum_{j=1}^{\NI/2} \left[  \frac{1}{|2j |} + \frac{1}{\sqrt{4 \left( \frac \Delta d \right)^2 + (2j-1)^2}} \right]  \right\}.
}
The term on the right-hand side of the first line expresses the contribution of the trapping potential, while, in the zigzag regime, the first term in the second line is the repulsion between ions in the same leg (upper or lower), whereas the second term corresponds to the repulsion between different branches. The classical equilibrium position $\Delta_0$ is determined by the minimum of $V(\Delta)$ or, equivalently, by the largest root of the derivative
\spl{
0 &\stackrel{!}{=} \frac 1 {\NI\;E_d } \frac{d}{d \tilde \Delta} V(\tilde \Delta)\\
&=2 \tilde \Delta    \left[ 1- \kappa \sum_{j=1}^\infty \frac{1}{[{4 \tilde \Delta ^2 + (2j-1)^2 }]^{3/2}} \right]
}
where $\tilde \Delta=\frac \Delta d$ is the dimensionless zigzag displacement.

Note that the overall electrostatic energy per ion diverges in the thermodynamic limit $\NI \to \infty$ at fixed $d$, i.e. the energy is not an extensive quantity (but super-extensive) due to the long-range nature of the Coulomb potential.

In this work we will not include the effect of the quantum fluctuations on $\Delta$, although the formalism provides a systematic starting point to consider them within a Hartree-Fock Bogoliubov-de Gennes approach. Both the equilibrium value of $\Delta$, as well as the transition point $\kappa_c$ are only modified slightly by the full quantum theory \cite{Silvi2013,Silvi2014}, as relevance of the quantum corrections is determined by the ratio of the local harmonic oscillator length scale and the distance between the ions $d$, which is very small in current experiments.

\section{Linear Ion Chain}\label{SEC:linear_chain}     

In this case the energy is minimized by $\Delta =0$ and it follows that $\overline \bR_l=l d \be_x$. In the translationally invariant case, the phonon frequencies and associated operators of the ion system can be determined analytically \cite{Fishman2008}, as we will show in the following within our formalism. The system features the highest order of discrete translational symmetry and the unit cell can be chosen to contain a single ion. The quasi-momentum $k$ takes discrete values, which can be restricted to the first Brillouin zone $k\in [-\frac \pi d, \frac \pi d)$ and equidistantly spaced by $\Delta k=\frac{2\pi}{\NI d }$.

The local, bare harmonic oscillator frequencies \refe{potential_expansion_coeff} can be evaluated in each dimension by summation of the infinite series in the thermodynamic limit. Although the energy per ion at fixed $d$ diverges, the local harmonic oscillator frequency converges to
\al{
\label{EQ:bare_ho_freq_x}
\Omega_{x}&=\sqrt{\frac{V_{(l,x),(l,x)}}{m_I}}= \sqrt{\frac{q^2}{4 \pi \epsilon_0 } \frac{4}{m_I d^3} \,\zeta(3) }\nonumber \\
& =\omega_I \sqrt{2 \kappa \zeta(3)}  \\
\Omega_{y/z}&=\sqrt{\frac{V_{(l,y/z),(l,y/z)}}{m_I}}=\sqrt{\alpha_{y/z} \omega_I^2 - \frac{q^2}{4 \pi \epsilon_0 } \frac{2}{m_I d^3} \,\zeta(3) }  \nonumber \\
&= \omega_I \sqrt{\alpha_{y/z}- \kappa \zeta(3)}  
\label{EQ:bare_ho_freq_yz}
}
where $\zeta(x)=\sum_{n=1}^\infty \frac{1}{n^x}$ is the Riemann zeta function (specifically $\zeta(3)\approx 1.2021$). We also defined $\alpha_y=1$ and $\alpha_z:=\alpha$, where $\alpha$ is again the anisotropy of the radial confinement.

The expansion coefficients in \refe{potential_expansion_coeff} can be evaluated (see Appendix~\ref{APP:taylor_exp_ion_pot} for details) explicitly and take on the the form
\spl{
\label{EQ:pot_deriv_Xk_Xl}
V_{(l,x),(l',x)}&=\left. \frac{\partial^2  V}{\partial  \bR_{l,x} \, \partial  \bR_{l',x}} \right|_{\overline \bR}
\\&=\frac{q^2}{4 \pi \epsilon_0 } \frac{1}{d^3} \left[  4 \delta_{l,{l'}}\, \zeta(3) - 2 \frac{(1-\delta_{l,{l'}})}{|l-{l'}|^3}\right]
}
\spl{
\label{EQ:pot_deriv_Yk_Yl}
&V_{(l,y/z),(l',y/z)}=\left. \frac{\partial^2  V}{\partial  \bR_{l,y/z} \, \partial \bR_{l',y/z}} \right|_{\overline \bR}
\\&\; =\frac{q^2}{4 \pi \epsilon_0 } \frac{1}{d^3}  \Bigg[ -2 \delta_{l,{l'}}\, \zeta(3)  +  \frac{(1-\delta_{l,{l'}})}{|l-{l'}|^3}\Bigg]+ \delta_{l,{l'}}\, m_I \alpha_{y/z} \omega_I^2,
}
where the last terms in the brackets of Eqs.~(\ref{EQ:pot_deriv_Xk_Xl} - \ref{EQ:pot_deriv_Yk_Yl}) are understood to only contribute if $l \neq {l'}$.

In the linear regime all cross terms of second order vanish identically
\spl{
\label{EQ:second_deriv_ion_between}
\left. \frac{\partial^2  V}{\partial \bR_{l,x} \, \partial \bR_{l',y}} \right|_{\overline \bR}&=\left. \frac{\partial^2  V}{\partial \bR_{l,x} \, \partial \bR_{l',z}} \right|_{\overline \bR}=\left. \frac{\partial^2  V}{\partial \bR_{l,y} \, \partial \bR_{l',z} } \right|_{\overline \bR}=0,
}
i.e. the  $(3\NI) \times (3 \NI)$-dimensional matrices $g$ and $h$ of Eq.(\ref{EQ:H_phonon}) are reducible into $\NI \times \NI$ uncoupled sub-blocks with respect to the different dimensions.

Using Eq.~(\ref{EQ:H_I_second_order_in_b}) we can write $\mH^{(2)}$ in the form 
\spl{
\label{EQ:H2_linear_chain_RS}
\mH^{(2)} &= E_0 + \frac 1 2 \sum_{l,\md} \Omega_{\md}( b_{l,\md}^\dag  b_{l,\md}^{\phantom{\dag}} +   b_{l,\md}^{\phantom{\dag}} b_{l,\md}^\dag  )\\
&  + \frac 1 2 \sum_{\stackrel{(l,\md),(l',\md')}{(l,\md)\neq(l',\md')} } \tilde f_{\md,\md'}(\Dis(l-l'))  \Big[
b_{l,\md}^\dag b_{l',\md'} + b_{l,\md}  b_{l',\md'}^\dag \\
& + b_{l,\md} b_{l',\md'} + b_{l,\md}^\dag b_{l',\md'}^\dag \Big]. 
}   
where we have separated the \textit{diagonal} (first row) and \textit{off-diagonal} (second and third row) terms and the local operators absorb all local terms of the potential, as well as the kinetic energy. We have also defined the coupling elements 
\spl{
\tilde f_{\md,\md'}(\Dis(l-l')) = \frac{V_{(l,\md),(l',\md')}}{2 m_I \sqrt{\Omega_{\md} \Omega_{\md'}}},
}
which only depend on the distance $\Dis(l-l')$ between site $l$ and $l'$, which is defined on a system with PBCs as the shorter of the two possible paths \footnote{For a system with PBCs, using the distance $\Dis(l-l')$ is equivalent to the normal distance on a an infinite straight line in the thermodynamic limit.}. In Eq.(\ref{EQ:H2_linear_chain_RS}) we  separated the purely local (\textit{diagonal}) terms $b_{l,\md}^\dag  {b_{l,\md}}^{\phantom{\dag}}$ from the non-local (\textit{off-diagonal}), and absorbed the former into the local harmonic oscillators. For the latter sum it is useful to define $\tilde f_{\md,\md'}(0):=0$, such that the summation can be performed over all ${(l,\md),(l',\md')}$, whereas in the original sum in \refe{H_I_second_order_in_b}  diagonal terms were excluded from the summation.

We can now exploit the discrete translational symmetry of the system, which allows one to express $\mH^{(2)}$ in a largely decoupled form by expressing it in terms of quasi-momentum creation and annihilation operators, which we define as the Fourier transform 
\al{
\label{EQ:ak_def}
a_{k,\md} &= \frac{1}{\sqrt{N}} \sum_l e^{-i d k l} {b_{l,\md}}.
}
Expressed in these operators, the coefficient matrix consists of uncoupled $2 \times 2$ sub-blocks, each of which corresponds to a fixed quasi-momentum $k$ and dimension $\md$. Subsequently, one can treat each such sub-block individually and the resulting phonon operators will be composed of the ladder operators associated with the respective sub-block only. 

We now explicate the procedure shortly. For a ring of $\NI$ ions, the allowed values of the quasi-momentum can be parametrized by an integer $n \in \{ 0 , \ldots , N-1\}$ as 
\spl{
\label{EQ:quasi-momentum_parametrization}
k_n=-\frac \pi d +n \frac{2 \pi}{\NI d}.
}

\paragraph{Off-diagonal terms}
Each of the four terms in the off-diagonal part of Eq.~(\ref{EQ:H2_linear_chain_RS}) comes with the same prefactor and transforms similarly into the $k$-representation up to coupling $k \leftrightarrow -k$. For instance, the first term transforms as
\spl{
&\frac 1 2 \sum_{l,l',\md,\md'} \tilde f_{\md,\md'}(\Dis(l-l')) b_{l,\md}^\dag b_{l',\md'} \\
&= \frac 1 2 \sum_{k,k',\md,\md'} f_{k,k',\md,\md'} a_{k,\md}^\dag a_{k',\md'}^{\phantom \dag}
}
with the discrete Fourier transformed coefficients
\spl{
\label{EQ:def_f_m_md}
 f_{k,k',\md,\md'} := \frac{1}{N}\sum_{l,l'} e^{-i d(k l  - k' l' )} \tilde f_{\md,\md'}(\Dis(l-l')).
}
As a consequence of the discrete translational symmetry and the separability in the spatial dimensions for the linear chain, one has $f_{k,k',\md,\md'} = \delta_{k,k'}\, \delta_{\md,\md'} \, f_{k,k,\md,\md}$ and we define $f_\md(k):=f_{k,k,\md,\md}$. For a system with PBCs and finite $N$, the vanishing of the off-diagonal $f$'s for $n \neq n'$ can be proven by changing the summation variables $(n,n')$ to relative $n-n'$ distances and \textit{center of mass} $n+n'$, with the summation borders being adjusted accordingly and is explicated in detail in Appendix~\ref{APP:vanishing_off_diagonal}. Due to the reflectional symmetry of the Coulomb force, $f_\md(k)\in \mathbbm R$.

\paragraph{Diagonal terms}
The local, diagonal terms [which have also been treated separately in Eq.~(\ref{EQ:H_I_second_order_in_b})] simply transform as
\spl{
\sum_l {b_{l,\md}}^\dag {b_{l,\md}}^{\phantom{\dag}} = \sum_k {a_{k,\md}}^\dag {a_{k,\md}}^{\phantom{\dag}}
}
and hence
\spl{
\frac 1 2 \sum_{l,\md} \Omega_{\md}( b_{l,\md}^\dag  b_{l,\md}^{\phantom{\dag}} +   b_{l,\md}^{\phantom{\dag}} b_{l,\md}^\dag  ) =  \sum_{k,\md} \Omega_{\md} \left[ a_{k,\md}^\dag a_{k,\md}^{\phantom{\dag}} + \frac 1 2 \right].
}
The quadratic Hamiltonian (\ref{EQ:H2_linear_chain_RS}) then becomes
\spl{
\mHph &= E_0 + \frac 1 2 \sum_{k,\md} \Omega_{\md}( a_{k,\md}^{{\dag}} a_{k,\md}^{\phantom{\dag}}  + a_{k,\md}^{\phantom{\dag}} a_{k,\md}^{{\dag}}) +\frac 1 2 \sum_{k,\md} f_\md(k) \\
& \times \Big[    a_{k,\md}^\dag a_{k,\md}^{\phantom{\dag}} +  a_{k,\md}^{\phantom{\dag}}   a_{k,\md}^\dag +  a_{k,\md}^\dag a_{-k,\md}^\dag + a_{k,\md}^{\phantom{\dag}} a_{-k,\md}^{\phantom{\dag}} \Big] \\ 
&= E_0 + \frac 1 2 \sum_{k,\md} {\vvec{a_{k,\md}^{\phantom{\dag}}}{a_{k,\md}^{{\dag}}}}^\dag    H^{(k,\md)}  \vvec{a_{k,\md}^{\phantom{\dag}}}{a_{k,\md}^{{\dag}}}
}
with the coupling matrix in the sub-block $(k,\md)$
\spl{
\label{EQ:H_phonon_k_delta}
H^{(k,\md)}  =\matr{\Omega_\md + f_\md(k) }{ f_\md(k)  }{f_\md(k) } {\Omega_\md + f_\md(k)}.
}
For the linear ion chain, the couplings can be evaluated analytically in the limit $\NI \to \infty$ at fixed $d$. One can express the sum over $k$ in the Fourier transform in Eq.~(\ref{EQ:def_f_m_md}) as a sum over integer $n$ using Eq.~(\ref{EQ:quasi-momentum_parametrization})
\spl{
f_\md(k)= 
\sum_{\Delta l=1}^{\NI} e^{-i d k\Delta l}   \frac{V_{\Delta l,\md,0 ,\md}}{2 m_I \Omega_\md }. \label{EQ:f_nu_summed} 
}
To obtain Eq.~(\ref{EQ:f_nu_summed}), we changed the summation $\sum_{l,l'=1}^\NI  $ to a summation over relative $l_r=l-l'$ and common variable $l_s=l+l'$. Together with the symmetry property $D(l_r)=D(\NI-l_r)$ and the summand being independent of $l_s$, the second sum simply yields a multiplicative factor $\NI$. To evaluate the final form $f_\md(k)$, we evaluate the $x$ and $y/z$ dimensions individually. Inserting the expressions from \refe{pot_deriv_Xk_Xl} into \refe{f_nu_summed}, one finds for the axial direction
\spl{
f_x(k)&= - \frac{q^2}{4 \pi \epsilon_0 d^3 \mI \Omega_x } \sum_{l_r=1}^{\NI} \frac{\left(e^{-i d k}\right)^{l_r}}{l_r^3}\\
&=-\frac{\kappa \omega_I^2}{2 \Omega_x} \, \mbox{Li}_3(e^{-ik d}),
}
where for the last line, we have taken the limit $\NI \to \infty$ and identified the last term as the polylogarithm of basis $s$
\spl{
\mbox{Li}_s(x)=\sum_{k=1}^\infty \frac{x^k}{k^s}.
}
Note, that in contrast to the potential energy which diverges super-extensively in the limit $\NI \to \infty$ at fixed $d$, these coupling elements, being proportional to the force, do converge. For the transverse coupling elements, by inserting \refe{pot_deriv_Yk_Yl} into \refe{f_nu_summed}, one analogously obtains
\spl{
f_{y/z}(k)&= \frac{\kappa \omega_I^2}{4 \Omega_{y/z}} \, \mbox{Li}_3(e^{-ik d}).
}
To determine the phonon frequencies and eigenvectors, one diagonalizes $\Sigma H^{(k,\md)}$, where $H^{(k,\md)}$ is given in Eq.~(\ref{EQ:H_phonon_k_delta}). As anticipated, the eigenvalues, if non-zero, appear in pairs of opposite sign $\pm\omega_{k,\md} = \pm \sqrt{(\Omega+f)^2-f^2}$ in accordance with general properties discussed in \refs{Transformation}. 

In addition to the spectrum, explicit analytic expressions for the phonon operators can also be obtained for the linear ion chain in the large $\NI$ limit.
These are obtained by taking the $\Sigma$-product of the eigenvector and the vibron operators. The $\Sigma$-normalized eigenvectors of $\Sigma H$ with $H$ given in \refe{H_phonon_k_delta} in compact form are
\al{
\bx^{(k,\md)}= \vvec{u^{(k,\md)}}{-v^{(k,\md)}}  = \vvec{  \sqrt{\frac{\Omega_\md+f_\md(k)}{ 2 \omega_{k,\md} } + \frac 1 2  }  }{ \sqrt{\frac{\Omega_\md +f_\md(k)}{ 2 \omega_{k,\md} } - \frac 1 2  }  }, \label{EQ:x_vector_elements}
}
as it can be directly verified. Together with \refe{def_beta}, \refe{x_vector_elements} leads to the explicit form of the phonon operators in terms of the microscopic position and momentum operators.

\subsection{Axial phonons}
Specifically, inserting $f_{x}(k)$ for axial direction, one finds the closed expression for longitudinal dispersion relation
\spl{
\omega_{k,x}&=\frac{q}{\sqrt{\pi \epsilon_0 m_I d^3}}  \sqrt{\zeta(3)-\mbox{Re}\left[ \mbox{Li}_3\left(e^{-ikd}\right)\right]}\\
&= \omega_I \sqrt{ 2 \kappa \left\{\zeta(3)-\mbox{Re}\left[ \mbox{Li}_3\left(e^{-ikd}\right)\right] \right\} }. \label{EQ:omega_eig_ax}
}
This expression, which is valid in the limit of a large system, contains no approximations and gives a linear, gapless dispersion relation at small $k$. The latter follows from the relation between the polylogarithm and the Riemann zeta 
\spl{
\mbox{Li}_s(1)=\zeta(s).
}

\subsection{Transverse phonons}

For the transverse phonons, on the other hand, the dispersion relation is given by
\spl{
\omega_{k,y/z}&=\sqrt{\omega_I^2 -   \frac{q^2}{2\pi \epsilon_0 m_I d^3}  \left\{  \zeta(3) - \mbox{Re}\left[\mbox{Li}_3\left(e^{-ikd} \right) \right]\right\}} \\
&=\omega_I \sqrt{1 -  \kappa \left\{  \zeta(3) - \mbox{Re}\left[\mbox{Li}_3\left( e^{-ikd} \right) \right]\right\}} \label{EQ:omega_eig_trans}
}
with $\mbox{Li}_3(-1) \approx -0.90154$.

Together, Eqs.~(\ref{EQ:Bog_op_trafos_matrix}, \ref{EQ:x_vector_elements}, \ref{EQ:omega_eig_ax}, \ref{EQ:omega_eig_trans}) constitute the full analytical form of the phonon operators for the longitudinal and transverse branches, expressed as a function of the microscopic parameters and the original position and momentum operators of each ion.

\subsection{Zigzag instability}
Since the ions interact repulsively, the one-dimensional chain is unstable towards transverse fluctuations of the ion positions depending on the strength of the transverse trapping potential. There is a critical transverse trapping depth, for which the system becomes stable, which depends on the inter-ion distance $d$.
Two criteria have to be fulfilled for the linear chain to be in stable equilibrium:
\begin{enumerate}
	\item All the bare local harmonic oscillator frequencies $\Omega_\md$, given in  Eqs.~(\ref{EQ:bare_ho_freq_x}) and (\ref{EQ:bare_ho_freq_yz}) have to be real, which sets an upper critical bound $\kappa < \tilde \kappa_c=\frac{1}{\zeta(3)} \approx 0.832$.
	\item All eigenvalues of $\Sigma \Hph$ are real and non-negative. This imposes 
	\spl{
\kappa < \kappa_c = \frac{1}{\zeta(3)-\mbox{Li}_3(-1) }=\frac{4}{7 \zeta(3)}\approx 0.4754,
}
since the instability always first appears at $k=\frac \pi d$ at the edge of the Brillouin zone, where $e^{-ikd}=-1$. The second criterion imposes a stronger restriction than the first by a factor of $4/7$, i.e. there exists a regime where the bare local harmonic oscillator frequencies are positive, but the system is nevertheless unstable towards zigzag formation. This underlines that the transition is a collective effect which cannot be understood from the analysis of a single ion in the effective potential of the other ions. Note that the above values for the transition are mean-field and do not incorporate quantum fluctuations. A full quantum analysis using Monte Carlo and other methods was performed in \cite{Astrakharchik2008, Silvi2013, Silvi2014}, where a small renormalization of the critical point was found when inserting typical experimental parameters. Using typical experimental parameters $d=5 \mu m$ and $\mI=10u$ (atomic mass units), one finds a renormalization of $\kappa_c$ by approximately $4 \times 10^{-4}$ when accounting for the quantum fluctuations. For larger $d$ or $\mI$ the renormalization of $\kappa_c$ is smaller.
\end{enumerate}
The second criterion is stronger and determines the transition: as $\omega_I$ is lowered, the transverse phonon branch moves down in energy and simultaneously the band width increases. The phonon modes at the edge of the Brillouin zone are generally lowest in energy and will reach zero energy first (this cannot be gauged away, as this energy corresponds to the difference in energy of the many-body eigenstates and is invariant under a shift of the many-body energy). As the mode branch touches zero, the mode frequencies vanish and thereafter become imaginary if the expansion is performed around the linear chain configuration.

\section{Phonon Spectrum in the Zigzag Phase}\label{SEC:zigzag_chain}
In this section, we derive the phonon spectrum in the zigzag configuration of an ion string and analyze the properties of the modes in detail. The mathematical structure is very similar to that of the linear chain discussed in \refs{linear_chain}, although the symmetry is reduced, leading to larger reducible sub-blocks in the coupling matrix and preventing an explicit analytic evaluation of all phonon frequencies and operators. The fluctuation expansion is performed around a classical zigzag configuration, which breaks the local $O(2)$ symmetry of the initial Hamiltonian. This couples the axial motion to the radial motion of the plane in which the zigzag transition occurs. Since the system can minimize its energy by choosing a non-zero $\Delta$, the expansion of the (operator) potential is performed around this configuration.
The unit cell is thereby doubled, now containing two ions (six degrees of freedom) with the Brillouin zone being reduced to $k \in \left[-\frac{\pi}{2 d}, \, \frac{\pi}{2 d} \right) = \mbox{1BZ}$.

\pic{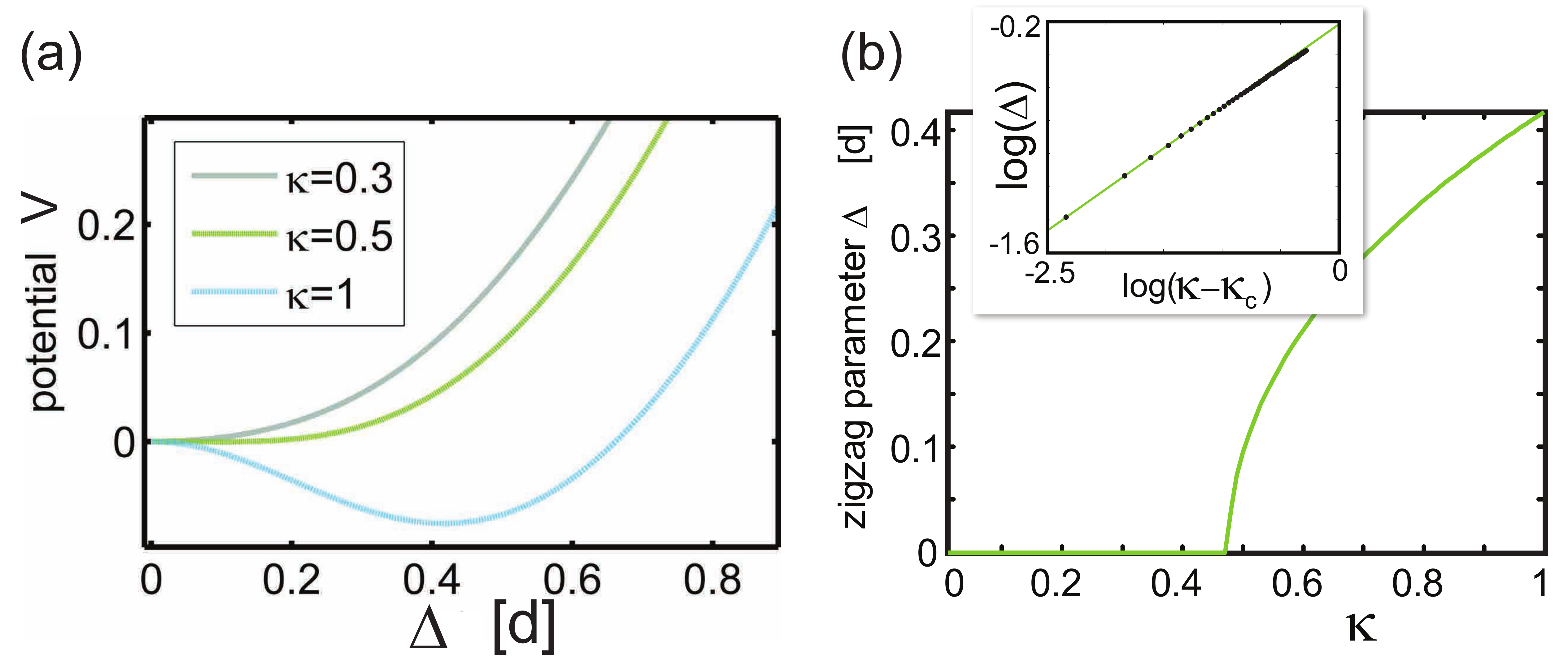}{(a) \label{FIG:V_of_Delta_and_Delta_of_kappa} The potential $V(\Delta)$ due to the Coulomb interaction and the external trap [see \refe{potentialdelta}] as a function of the zigzag parameter $\Delta$ (the coordinate at $k=\pi/d$) for various values of the dimensionless effective Coulomb interaction strength $\kappa$. In the linear regime the global minimum $\Delta_0$ is at $\Delta=0$, whereas in the zigzag regime $\Delta_0$ moves out to non-zero $\Delta$. For the axial symmetric case, the potential in the $y$,$z$ coordinates would be a mexican hat potential, with the system choosing a direction and spontaneously breaking the symmetry. (b) $\kappa$-dependence of the order parameter $\Delta_0$, defined by the value of $\Delta$ which minimizes the potential in (a). Inset: the black dots show the same function, but on a log-log scale. The green line shows $\sqrt{ \kappa-\kappa_c }$ for comparison, demonstrating the expected mean-field critical behavior.}{\linewidth}

To account for the doubling of the unit cell, (a symmetry reduction relative to the linear chain), we transform into the intermediate basis and define the ladder operators
\spl{
a_{k,s,\md}^\dag:=\sqrt{\frac{2}{\NI}} \sum_{l=1}^{N/2} e^{i\, 2d\, kl} b_{2l+s,\md}^\dag
}
where $s\in\{0, \, 1\}$ denotes the sublattice (upper or lower row of the zigzag) on which the intermediate mode lives and $\md \in \{x,y,z\}$ is the direction. The inverse relation is given by
\spl{
\label{EQ:def_b_op_zigzag}
b_{2l+s,\md}^\dag= \sqrt{\frac{2}{\NI}} \sum_{k \in \mbox{\tiny 1BZ}} e^{-i\, 2d\, kl} a_{k,s,\md}^\dag.
}

In \reff{V_of_Delta_and_Delta_of_kappa}, the behavior of the zigzag order parameter $\Delta_0$ is shown for various $\kappa$. If $\kappa$ is increased beyond $\kappa_c$, the equilibrium position of $\Delta$ around which the expansion of the potential has to be performed moves to a non-zero $\Delta_0>0$. Close to the transition the mean-field behavior of the zigzag order parameter is demonstrated in the inset of \reff{V_of_Delta_and_Delta_of_kappa} (b). An explicit expression for $\Delta(\kappa)$ at arbitrary values of $\kappa$ is not known to the authors and, although the derivative of the potential can be calculated, the coefficients $V_{(j,\md),(j',\md')}$ at the minimum have to be evaluated numerically. 

Given the $V_{(j,\md),(j',\md')}$, the second order Hamiltonian is then of the form in \refe{H_I_second_order_in_b}, which we express in terms of the symmetry-adapted annihilation operators $a_{k,s,\md}$ defined in \refe{def_b_op_zigzag}. In contrast to the linear case, one now also has coupling between the modes on different sublattices $s$ and $s'$, as well as between modes in the $x$ and $y$ dimensions, but quasi-momentum in the reduced Brillouin zone is preserved. This leads to the coupling matrix being reducible into  sub-blocks of size $8 \times 8$ and $4 \times 4$ for each $k$.

Specifically, for each $k$ we define
\spl{
\label{EQ:ak_column_vecs}
\ba_{k}=\begin{pmatrix}  a_{k,0,x}^{\phantom{\dag}} \\ a_{k,1,x}^{\phantom{\dag}} \\ \alpha_{k,0,y}^{\phantom{\dag}} \\ a_{k,1,y}^{\phantom{\dag}} \\ a_{k,0,z}^{\phantom{\dag}} \\ a_{k,1,z}^{\phantom{\dag}} \end{pmatrix} , \quad 
\ba_{-k}^\dag=\begin{pmatrix}  a_{-k,0,x}^\dag \\ a_{-k,1,x}^\dag \\ a_{-k,0,y}^\dag \\ a_{-k,1,y}^\dag \\ a_{-k,0,z}^\dag \\ a_{-k,1,z}^\dag \end{pmatrix}
}
(note that it is useful to define both $\ba_{k}$ and $\ba_{-k}^\dag$ as column vectors in this case), such that the second order Hamiltonian can be expressed as
\spl{
\mHph=  E_0+ \frac 1 2 \sum_k  {\vvec{\ba_{k}}{\ba_{-k}^\dag}}^\dag    \Hph^{(k)}  {\vvec{\ba_{k}}{\ba_{-k}^\dag}}.
}
with the coupling matrix in the sub-block $k$
\spl{
h^{(k)}, \, g^{(k)}=
\begin{pmatrix} 
\begin{pmatrix} 
*& * & * & * \\
*& * & * & * \\
*& * & * & * \\
*& * & * & * 
\end{pmatrix}
 & 0 \\ 0 & 
\begin{pmatrix} 
*& *  \\
*& *  
\end{pmatrix}
\end{pmatrix}
}
\spl{
\Hph^{(k)}=
\begin{pmatrix} 
h^{(k)} & g^{(k)} \\
g^{(k)^*} & h^{(k)^*}\\
\end{pmatrix}.
}

As reflected by the structure of $\Hph^{(k)}$, the symmetry of the potential ensures that the helical excitations in the $y$-direction are decoupled from the excitations in the $x$-$z$-plane, allowing for a further reduction and a separate diagonalization of a $8\times 8$ and a $4\times 4$-dimensional matrices. Note that for each $k\neq 0$, there are two independent and uncoupled sub-blocks $\Hph^{(k)}$ and $\Hph^{(-k)}$. For systems with time reversal symmetry, the elements can be chosen to be identical. For $k=0$ there is only one such block, such that particle and hole excitations of the same original $\alpha_{k=0}$ mode are coupled.

The diagonalization of $12\times 12$-dimensional $\Sigma \Hph^{(k)} $ at each $k$ is conceptually identical to the procedure for the linear chain. However, with the sublattice-dimension index pair $(s, \nu)$ entering the transformation at every $k$, each emerging mode is characterized by internal mode label $\gamma \in \{ 1, \ldots, 6 \}$ together with its quasi-momentum. In the dispersion relations shown in \reff{dispersion_rels_combined_V3}{}, $\gamma$ labels one of the six modes at each $k$. In both the zigzag and the linear regimes, each mode continuously crosses over into another branch (which is folded back into the first Brillouin zone) without a band gap opening. Each physical phonon mode branch can thus be understood to consist of two branches, which cross over into each other continuously at the edge of the Brillouin zone. For any given $k\neq 0$, one thus finds six independent eigenvectors with a positive $\Sigma$-norm  $\bx^{(k,\gamma)}=\vvec{\bu^{(k,\gamma)}}{-\bv^{(k,\gamma)}}$. The lower indexes of $u_{s,\md}^{(k,\gamma)}$ of each eigenvector can be understood as a collectiev index for the element, whereas $(k,\gamma)$ labels the eigenvector. In the zigzag regime at $k=0$, one only finds four positive eigenvectors $\bx$ and two linearly independent vectors $\bp^{l,r}$ with a zero $\Sigma$-norm, corresponding to the two gapless modes in both longitudinal and radial direction. The Hamiltonian to second order is given by
\spl{
\label{EQ:H2_diagonal_zigzag}
\mHph =&E_0+\Delta E^{(2)} +\sum_{k,\gamma}  \omega_{k,\gamma} \beta_{k,\gamma}^\dag \beta_{k,\gamma} + \frac{\mathscr P_l^2}{2 \tilde m_l} + \frac{\mathscr P_r^2}{2 \tilde m_r}.
}
Here, $\Delta E^{(2)}=\frac 1 2 \sum_{k,\gamma} \omega_{k,\gamma} $ is the zero point energy shift analogous to \refe{zero_pt_en_shift}. For completeness, we here included both effective free particle degrees of freedom in the zigzag regime with $\tilde m_l$ and $\tilde m_r$ referring to the longitudinal and radial mass-like terms respectively. In the linear regime, the last $\mathscr P_r$ term falls away and is replaced by a true phonon mode at $k=0$ with a non-zero frequency. Also, the results from the previous section have to be reproduced by this treatment in a two ion unit cell, each emerging mode index $\gamma$ can be decomposed into a dimensional index (being a good quantum number of the excitation in the linear ion chain) and an additional index specifying the branch number, as the dispersion relation is folded back into new first Brillouin zone at the edge.

In the zigzag regime, there is one helical twisting branch, which decouples from the dynamics within the zigzag plane and is characterized by the motion of the ions out of the zigzag plane. Being the gapless mode (analogous to a Goldstone mode in 3D) associated with the spontaneous breaking of the $O(2)$, this mode is gapless and linear in the lower branch at small $k$ and is shown in light blue in \reff{dispersion_rels_combined_V3} (a3, a4). The other two modes are associated with the ion motion within the zigzag plane, but since the operators between the radial and longitudinal degrees of freedom are now coupled by a non-zero zigzag parameter $\Delta$, the resulting modes are now hybridized, i.e. no longer correspond to motion in pure radial or longitudinal direction. In fact, the direction of the motion is now $k$-dependent and the direction of oscillation changes continuously with $k$, as characterized by the spatial mixing angle (defined and discussed in the following section).

\pic{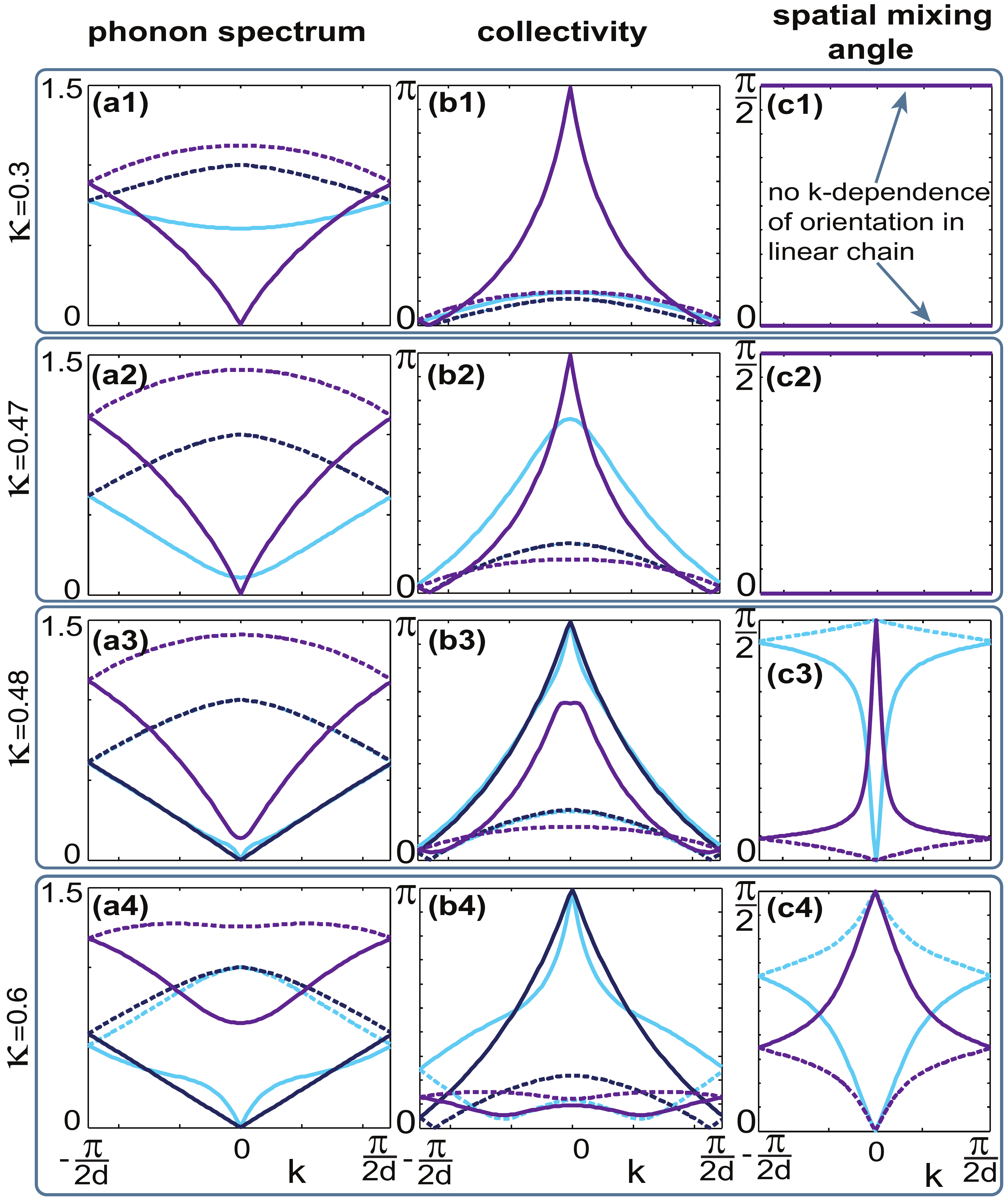}{\label{FIG:dispersion_rels_combined_V3} The dispersion relations [column (a)], collectivity [column (b)] and spatial mixing angle [column (c)], each shown as a function of the quasi-momentum $k$. The rows correspond to different values of the interaction $\kappa$, with the upper and lower two rows corresponding to the linear ($\kappa< \kappa_c$) and zigzag ($\kappa> \kappa_c$) regimes respectively. Identical line color and style throughout the different plots correspond to the same mode branch and, when crossing the transition, matching the modes with the largest overlap. Throughout the entire parameter range, the set of all modes (six at every $k$) form three closed loops, i.e. one loop containing two phonon modes at every $k$ and converging to the same parameter (either frequency, collectivity or spatial mixing angle) at $k \to \pm \pi/2d$.}{\linewidth}

The resulting spectra across the zigzag transition (also calculated by Fishman et al. \cite{Fishman2008}) are depicted in the left column of \reff{dispersion_rels_combined_V3}. As $\kappa$ is increased in the linear regime, the two radial modes move down in frequency, an effect which is pronounced at small $k$. At the zigzag transition point, the two radial modes at $k=0$ reach zero energy, which means that it costs no energy to populate these phonon modes. 
This can be understood as a condensation of phonons into the zigzag mode. If the condensed state is taken as a coherent state in the $k=\pi/d$ mode (corresponding to a superposition of states of different phonon number), this corresponds to the most classical configuration of ions in the spatial zigzag configuration. Hence, even in the zigzag regime the entire calculation could still be performed in the basis of phonons of the linear chain with the shifted ground state being a phonon condensate, but terms beyond the quadratic expansion are required to limit the zigzag amplitude ($\propto$ phonon number per ion) to a finite value.  

In addition to the axial gapless sound mode, a further gapless mode associated with the spontaneous $O(2)$ symmetry breaking in the isotropic system appears on the zigzag side of the transition $\kappa>\kappa_c$. This is associated with a helical flexing motion of the ions, the frequency of which decreases to zero in the long wavelength limit. At $k=0$ the entire zigzag chain rotates collectively around the radial axis, a motion which has no restoring force and therefore cannot be associated with a bosonic mode, but resembles that of an effective free particle degree of freedom with PBCs, as reflected by the last term in \refe{H2_diagonal}. Note that in the zigzag regime there are two independent effective momentum  $\mathscr P_{r/l}$ and position $\mathcal Q_{r/l}$ operators for the two gapless modes. Since the different $\mathscr P_{r/l}$ operators are associated with degenerate eigenvalue zero, these have to be orthogonalized or associated by symmetry, e.g. with the longitudinal and axial motion.

\paragraph{Spatial Mixing Angle}
The modes (in the sense of the eigenvectors of $\Sigma \Hph^{(k)}$ in an appropriate gauge) evolve smoothly with $k$ and also do not split at the edges of the Brillouin zone. Due to the coupling of the $x$-$y$ motion in the zigzag regime, the dimensional index is no longer a good quantum number. In fact, the spatial orientation of the mode's motion changes continuously with $k$, which we characterize by defining the \textit{spatial mixing angle} for a mode $(k,\gamma)$
\spl{
\theta_{xy}(k,\gamma)=\arctan \left[ \frac{ \sum_{s} \left(|u_{y,s}^{(k,\gamma)}|^2 + |v_{y,s}^{(k,\gamma)}|^2 \right) }{  \sum_{s} \left(|u_{x,s}^{(k,\gamma)}|^2 + |v_{x,s}^{(k,\gamma)}|^2 \right)  } \right].  \label{EQ:mixing_angle}
}
The weights of both the particle and the hole excitations is taken into account (with respect to the bare local harmonic oscillator levels) by this definition and it quantifies the motional orientation as an angle of an ion when the system is excited in the respective mode. This is shown for various values of $\kappa$ in the right column of \reff{dispersion_rels_combined_V3}. In the linear regime, there is no mixing of the modes and the values of $\theta_{xy}(k)$ are pinned to $0$ or $\pi/2$. As soon as the zigzag regime is entered, the values at $k=0$ points remain pinned to either of these values (a consequence of the symmetry of the potential and the time reversal symmetric point), but at opposite ends of one closed loop. This behavior is reflected in the hypothetical dynamics of phonon Bloch oscillations in the zigzag regime: suppose we prepare a phonon state in a fixed $k$ mode, e.g. a phonon Fock state $\ket{k,n} = \frac{(\beta_k^\dag)^n}{\sqrt{n!}} \ket{\psi_0}$ or a coherent state $\ket{\psi_k,n} = e^{\sqrt{n}(\beta_{k,\gamma}^\dag-\beta_{k,\gamma})} \ket{\psi_0}$, where $\ket{\psi_0}$ is the phonon ground state and we then apply a constant effective force to the phonons (analogous to a tilted potential in a localized phonon basis), the excitation will traverse this closed trajectory and continuously change its spatial orientation of motion. Every time it passes $k=0$, it will have flipped its orientation by $\frac \pi 2$ in the zigzag plane. At small $\kappa-\kappa_c$, $\theta_{xy}(k)$ changes rapidly with $k$ in the vicinity of $k=0$, but this rate slows down at larger $\kappa$. It is also worth noting that for the four given modes in the zigzag plane, the spatial mixing angles of two modes $(k,\gamma)$ and $(k,\gamma')$ fulfill $\theta_{xy}^{(\gamma)}(k)+ \theta_{xy}^{(\gamma')}(k) = \pi/2$ throughout the entire range of $k$. 

\paragraph{Collectivity}
Another quantity that is useful in quantifying and characterizing the phonon mode structure is the collectivity, which we define as $\mC_s(k)= \frac{|\bv^{(k)}| }{|\bu^{(k)}|}$ for each phonon. With this definition, $\mC_s(k)$ can vary between zero and one. For $\mC_s(k)=0$ (implying $|\bv^{(k)}|=0$) the phonon excitation is fully composed of excitations of the bare local phonon states and in this sense `particle-like'. In the opposite limit $\mC_s(k)=1$, a phonon excitation is an equal, hybridized mixture of particle and hole excitation in the bare local phonon basis.

The name is chosen in analogy to the characterization of excitations in condensed matter, where  `particle-like' excitations can be understood in terms of the excitation structure of a single particle above, whereas if both $|\bu^{(k)}|$ and $|\bv^{(k)}|$ are large, such an interpretation is impossible and has to be understood in a collective context, involving all particles. When thinking of phonon modes in trapped ion systems as superpositions of excitations of the bare local harmonic oscillator states, the collectivity quantifies the validity of this picture: if $\mC_s(k)$ is large, one can think of a single excited phonon as coherent excitations, consisting both of excitations above and anti-excitations below a reference state. In other words, $\mC_s(k)$ is a measure for the number of virtually populated bare local oscillators in the phonon ground state $\ket{\psi_0}$.

We show the collectivity as a function of $k$ for various $\kappa$ in the central column of \reff{dispersion_rels_combined_V3}. Typically, the collectivity of the excitation increases as the mode lowers its energy, as can be seen in the $k \to 0$ limit for the gapless longitudinal and helical modes. Interestingly, this is not true in general though. In fact, in certain regimes the collectivity may actually decrease and feature non-monotonic behavior with $k$, as can be seen for the lower branch of the radial gapped mode in \reff{dispersion_rels_combined_V3} (b3-b4).

\section{Effective Free Particle Degree of Freedom}\label{SEC:zeromode}
In Sec.~\ref{SEC:PhononTransformation}C we showed that in the case of gapless modes, not all operators can be expressed in terms of phonon operators, but an additional effective free particle degree of freedom also arises. Specifically, this is the case in both the linear and the zigzag chain in the axial direction, which can be intuitively understood, as there is no restoring force if all ions move in the same manner. If the ions would be trapped in an axial ring shaped structure (realizing periodic boundary conditions), such a system would continue rotating in the absence of dissipative coupling and higher order terms $\mH_\Delta$. While this setup may be hard to realize in current experiments, the analogous phenomenon arises naturally in the radial degrees of freedom in the zigzag regime with radial rotational symmetry $\alpha=1$. In this case, there is also no restoring force if the entire zigzag chain is set into rotational motion it will perpetually rotate, following the dynamics of an effective free particle \textit{living} on a one-dimensional space with PBCs.

We shall now systematically construct suitable Hermitian operators with the position and momentum, which incorporate the PBCs. However, we first note that up to now, the ion system was completely specified by the single dimensionless parameter $\kappa$: for a given value of $\kappa$, all quantities, such as the dispersion relations given in units of $\omega_I$ were universal functions. This implies a scale invariance of the system when neglecting the effects of quantum fluctuations at the transition point. However, when including the effects of the helical effective free particle degree of freedom, specifying only $\kappa$ is no longer sufficient to fully characterize the system and an additional parameter is required. Within the trapped ion chain, one actually has two natural length scales (as well as associated energy scales) and the additional dimensionless parameter can be thought of as the ratio of these.
For the length scales, one has the ion spacing $d$, as well as the harmonic oscillator length $(\mI \omega_I)^{1/2}$ of an ion associated with the radial trapping frequency $\omega_I/(2\pi)$ (the exact bare trapping frequencies in the different dimensions are renormalized by the effect of the other ions and related to $\omega_I$ by scalar functions of $\kappa$ of order one). We define the ratio of these length scales
\spl{
\lambda:= d \sqrt{\mI \omega_I}.
}
The two energy scales $E_d= \mI \omega_I^2 d^2 / 2$ and $\omega_I$ can then also be expressed in terms of $\lambda$ as ${E_d}/{\omega_I}=  \lambda^2 / 2$. This parameter becomes relevant in the zigzag regime. The natural length scale for the zigzag amplitude $\Delta_0 d$ is $d$, whereas the intrinsic length scale of the ion wave packet is set by ${E_d}/{\omega_I}$. For small $\Delta_0$, if these tow length scales are of the same order, the radial particle dynamics is always affected by the PBCs and the discrete structure of this subsystem's energy spectrum (imposed by the PBCs) is relevant to the dynamics. The formalism we present here applies to both the helical and axial zero energy degrees of freedom, with the difference that the length of the system (with PBCs) is $2\pi \Delta_0 d$ and $\NI d$ respectively. Let us refer to this circumference as $L$ for the discussion here. Similarly, the local harmonic oscillator frequency $\Omega$ refers to $\Omega_x$ and $\Omega_z$ respectively.

The normal form of the linearized equations of motions inherently predicts the effective free particle degrees of freedom in \refe{H2_diagonal_zigzag}, which cannot be expressed in terms of phonon operators, but in terms of conjugate operators $\mathscr P$ and $\mQ$. Also, the reciprocity of these operators is set by their commutation relations. What is not predicted correctly, however, is the space on which they act, since they were obtained from a linearization around a localized, classical zigzag configurations - an expansion which is local in character and cannot be aware of the global topology and the PBCs. To unite these two approaches, i.e. 1) the local existence of $\mathscr P$ and $\mQ$ and $[\mQ, \mathscr P]=i$ together with 2) the periodic nature of $\mQ$ on the full space, we construct a Hermitian position (angular phase) operator $\hat \varphi$ from the energy eigenstates respecting the PBCs (see Appendix~\ref{SEC:PBC_position_op} for details). Note that from a mathematical perspective, this construction features strong similarities to the construction of the Pegg-Barnett phase operator \cite{Barnett1986}. As in the latter case, the usual commutation relation $[\mQ,\mathscr P]=i$ can no longer be fulfilled in the strict sense at a global level when incorporating the PBCs. The diagonal elements of $\bra m \mathcal Q\ket m=0$ in the energy eigenbasis representation vanish. Both operators $\mQ$ and $\hat \varphi$ are dimensionless. Matching their local properties implies that the extension of $\mQ$ to the global space has to be a scalar multiple $\mQ=c_0 \hat \varphi$ with a dimensionless $c_0$ that we shall now determine.

For brevity we focus on the radial effective free particle degree of freedom. Since $\mQ$ in \refe{Q_def} is dimensionless and the natural units of $\delta \bR_{j,s,\md}$ in the previous analysis is $[{\mI \Omega_{\md}}]^{-1}$ together with the form of $\bq$ leads to 
\spl{
\mQ=\frac 1 N \sum_{j,s}(-1)^s \sqrt{{\mI \Omega_z}} \; \delta \bR_{j,s,\md=z}.
}
Hence, if the entire zigzag plane undergoes one full rotation, the eigenvalues of any given $\delta \bR_{j,s,\md=z}$ range over $[0, \; 2\pi \Delta_0 d)$, the eigenvalues of $\mQ$ have to range over
\spl{
\mbox{range}(\mQ)&= [0, \; 2\pi \Delta_0 d \sqrt{{\mI \Omega_z}} )\\
&= [0, \; 2\pi \Delta_0 \lambda \sqrt{\frac{\Omega_z}{\omega_I}} ).
}
This fixes $c_0=\Delta_0 \lambda \sqrt{\frac{\Omega_z}{\omega_I}}$ and thus
\spl{
\mQ=\Delta_0 \lambda \sqrt{\frac{\Omega_z}{\omega_I}} \hat \varphi.
}
Let $\hat \Pi$ be the conjugate operator to $\hat \varphi$, which acts on any energy eigenstate $\ket m$ as $\hat \Pi \ket m = m \ket m$. Then, by reciprocity, $\mathscr P= c_0^{-1} \hat \Pi$, allowing us to consequently evaluate the physical energy in \refe{H2_diagonal} of any eigenstate $\ket m$ as
\spl{
\label{EQ:Psq_ev}
E_m^{(r)}:=\bra m\frac{\mathscr P^2}{2 \tilde m_r} \ket m= \frac{1}{2 \tilde m_r} \frac{\omega_I \, m^2}{\lambda^2 \Delta_0^2 \Omega_z} 
}
and we remark that $\tilde m_r$ has the dimension of an inverse energy. The thermal expectation value (working in units $k_B=1$)
\spl{
\ev{\mathscr P^2}= \frac{\sum_m e^{-E_m^{(r)}/T } \bra m \mathscr P^2 \ket m }{\sum_m e^{-E_m^{(r)}/T } }
}
is shown in \reff{P_sq_eval} for various temperatures $T$. At $\kappa > \kappa_c$, where the level spacing becomes thermally accessible, the higher $\ket m$ states become significantly occupied and can absorb a significant amount of entropy.

\pic{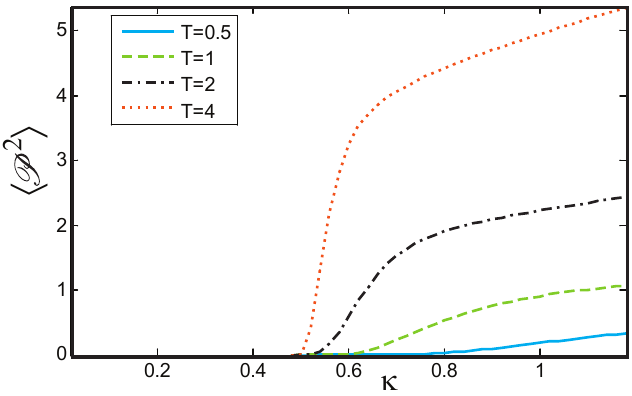}{\label{FIG:P_sq_eval} The thermal expectation value $\ev{\mathscr P^2}$ for the radial effective free particle degree of freedom, associated with the rotation of the zigzag plane. Since the energy levels grow quadratically in the level number, the contribution grows strongly with increasing temperature $T$, as shown in the different lines corresponding to different $T$. This expectation also contributes to the total energy of the system in conjunction with $\tilde m_r$ as $\ev{\mathscr P^2}/(2\tilde m_r)$, where $\tilde m_r$ also depends on $\kappa$ (shown in \reff{m_tilde_of_kappa}). The various curves are shown for $\omega_I/(\lambda^2 \Omega_z)=1$.}{7cm}

Physically, each energy eigenstate (analogous to 2D angular momentum eigenstates) is fully delocalized on the periodic ring on which they live and $m$ corresponds to the phase winding number of each eigenstate. Evaluating the spatial variance leads to
\spl{
\label{EQ:ev_Qsq}
\bra m \mQ^2 \ket m = \frac{\pi^2 \Delta_0^2 \lambda^2 \Omega_z}{3 \omega_I},
}
independent of $m$ in the limit of a large cut-off for the states $\ket m$. This  result is independent of $m$ and in agreement with the naive variance ($\pi^2 / 3$ for a unit circle) of a flat distribution over the circumference up to the scaling factor $c_0^2$, as shown in \refs{PBC_position_op}. The independence of $m$ implies that this variance also holds for thermal expectation values at arbitrary temperature. Furthermore, it can be shown that
\al{
\label{EQ:ev_Q_PQ}
\bra{l} \mathcal Q \ket{l}&=\bra{l} \mathscr P \mathcal Q \ket{l}=0.
}
The above relations in Eqs.~(\ref{EQ:Psq_ev} - \ref{EQ:ev_Q_PQ}) are of importance for the evaluation of observables, as will be shown in \refs{spatial_correlations} for example. 
For the ground state of the Hamiltonian theory (without dissipation), the terms $\mathscr P^2$ in the Hamiltonian favors a delocalization of $\mathcal Q$, i.e. a macroscopic superposition of states with different orientations of the zigzag plane $\ket{\psi_0} = \int_0^{2\pi} \frac{d \varphi}{2\pi} \ket{\phi_0(\varphi)}$, where $\ket{\phi_0(\varphi)}$ is a phonon ground state with a fixed zigzag plane orientation $\varphi$ (see Appendix~\ref{SEC:PBC_position_op} for a relation with the $\mQ$ operator). However, one would expect such a state to be very susceptible to decoherence, as the charged ions couple strongly to small electric field fluctuations.

We point out that from a mathematical perspective, the concepts of states with fixed angular orientation $\ket{\phi_0(\varphi)}$ as counterparts to angular delocalized states such as $\ket{\psi_0}$ are analogous to states with fixed total particle numbers as counterparts to states with a well-defined phase (spontaneously breaking the $U(1)$ symmetry) in Bose-Einstein condensates or BCS superconductors \cite{Tinkham2004}. In both cases the term $\mathscr P^2$ in the Hamiltonian energetically penalizes a localization in the phase. However, whereas in systems such as atomic gases or superconductors the total particle number is often a strictly conserved quantity, there is no analogous restriction for the angular orientation of the ion chain.

\subsection{Effective mass-like constant}

There are potentially two effective free particle degrees of freedom for an infinite linear, or finite ring ion chain. In both the linear and zigzag phases where the axial translational symmetry is spontaneously broken, the collective center of mass motion along this direction has no restoring force and can be described by an effective free particle with mass $\tilde m_{l}$. Remarkably, the effective mass-like constant along the axial direction $\tilde m_{l}$ is not simply a constant multiple of the bare ion mass, but depends non-trivially on $\kappa$, as shown in \reff{m_tilde_of_kappa}. At the critical point $\kappa_c$, this effective mass $\tilde m_{l} \omega_I / N$ takes on a minimum and features a kink. In the non-interacting limit $\kappa \to 0$ and for the chosen units, $\tilde m_{l} \omega_I / N$ diverges. In contrast, the helical $\tilde m_r$ starts off at a finite value at $\kappa_c$ and increases steadily with $\kappa$.

\pic{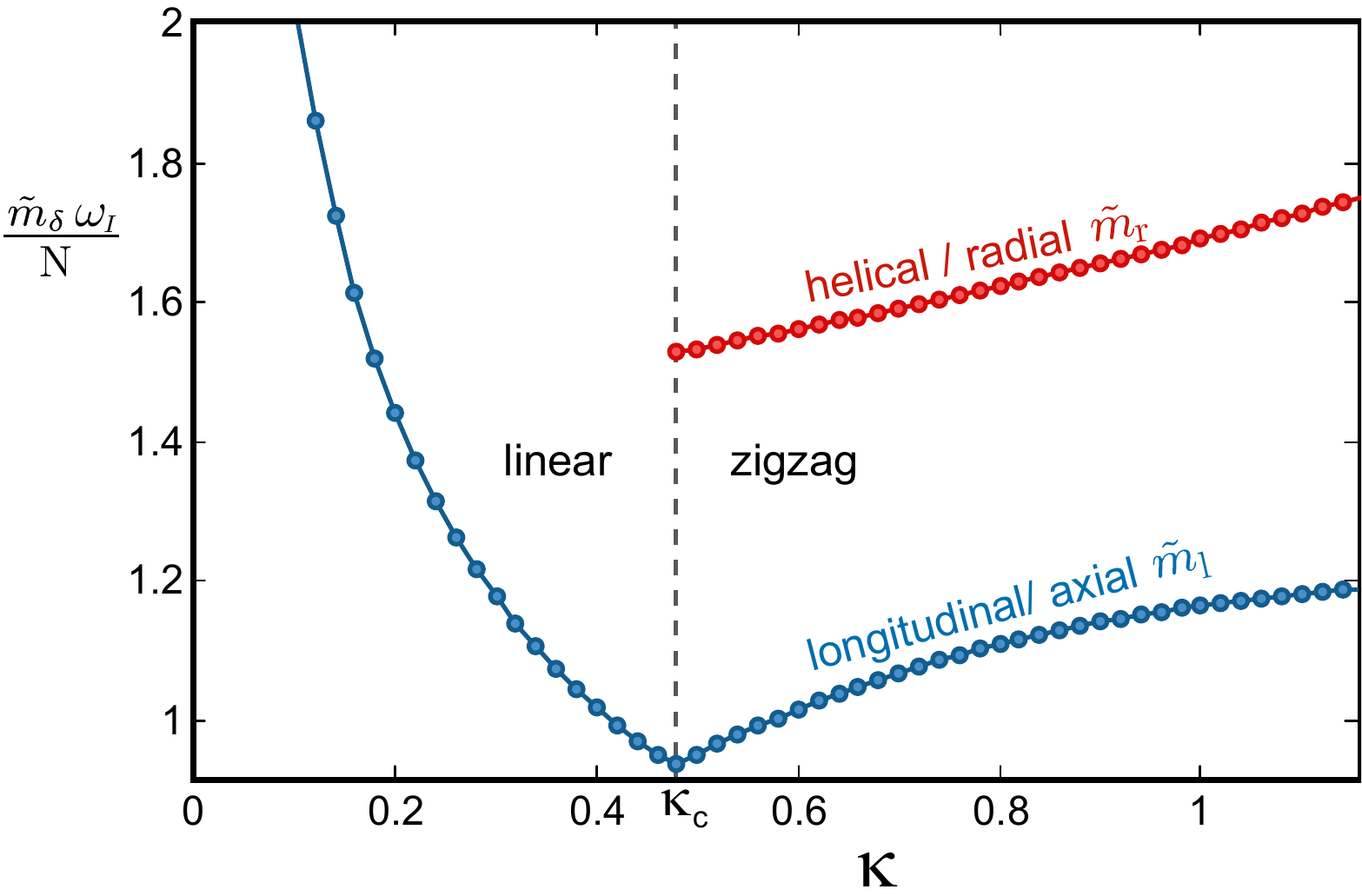}{\label{FIG:m_tilde_of_kappa} The effective mass-like constant $\tilde m_\delta$ (with $\delta=$r/l for the radial / longitudinal case respectively) associated with the free particle degree of motion in units of $\omega_I^{-1}$. In the linear ion chain for $\kappa < \kappa_c$ only a single effective free particle degree of freedom exists (associated with axial motion and shown in blue), whereas in the zigzag regime for $\kappa > \kappa_c$ an additional, independent effective free particle emerges (associated with a rotation of the entire zigzag plane, shown in red).}{7cm}

\section{Fluctuations and Ginzburg Criterion}
\label{SEC:Ginzburg_crit}
For a 1D system with spontaneously broken continuous symmetries, such as the 1D ion chain, the Mermin-Wagner theorem forbids the formation of strict 1D order in the thermodynamic limit. This naturally also applies to the ion chain and is relevant to observables for quantities such as the behavior of correlation functions above a certain length scale. The absence of true long range order for $\alpha=1$ is inherently reflected by our approach, e.g. by determining the Ginzburg parameter \cite{Amit1974} or the logarithmic divergence of correlation functions in the thermodynamic limit, as we will now shortly discuss.

The absence of true long range order applies to both the axial and helical modes (for the latter only in the zigzag phase). Since the former is typically irrelevant to experiments and the underlying principle is identical, we here focus on the absence of true long-range zigzag order. The latter is the underlying assumption, around which the expansion of small fluctuations is performed. Specifically, this implies that the spatial variance of any single ion along the helical direction (i.e. the $y$-direction to linear order in our convention) is small relative the diameter $2 \pi \Delta_0$, i.e. the Ginzburg parameter $\ev{\delta \bR_{j,s,\nu=y}^2}/(2 \pi \Delta_0)^2$ should be small.

Let us consider the numerator only and show that this diverges, i.e. that quantum fluctuations lead to a divergence in the variance of the spatial distribution associated with the motion in the direction of the gapless mode. Specifically, let us evaluate $\ev{\delta \bR_{j,s,\md=y}^2} \propto \sum_{k,\gamma} | v_{s,\md}^{(k,\gamma)} |^2$. For any gapless mode, the norm of $v$ (and hence the elements) scales as $k^{-1/2}$, i.e. $\lim_{k \to 0} |v_{s,\md}^{(k,\gamma)}|^2 k = \mbox{const}$. In the thermodynamic limit $L \to \infty$, the sum $\sum_k$ can be replaced by an integral $\ev{\delta \bR_{j,s,\nu=y}^2} \propto \int_{}^{} dk \, \frac 1 k$. The inverse lattice spacing sets an inherent UV-cutoff, however the integral diverges logarithmically in $L$ from the $k\to 0$ contribution, as the lower integral limit is $\propto L^{-1}$.

For finite systems, on the other hand, this approach should still describe the correlations correctly in a regime away from the critical point. The distance from the critical point $\kappa-\kappa_c$ will determine a systems size $L$ up to which the correlations are approximately described. We therefore, in the following section, show the correlation functions influenced by a gapless mode for a finite size system $L=50$, whereas the correlators not containing any contribution from a gapless mode are shown in the thermodynamic limit. 

\section{Spatial correlations}
\label{SEC:spatial_correlations}
 
To evaluate the correlations present in the motion between the different ions, we calculate the spatial correlations both in the phonon ground state $\ket{\psi_0guara}$ and at finite temperature. In contrast to the simple \emph{classical} ground state, where each ion is in an uncorrelated, local Gaussian state of its local quadratic minimum, the phonon ground state also contains correlations at $T=0$. To leading order, and for typical ion experiments to an excellent degree of accuracy, the leading contribution beyond the classical, scalar contribution is given by the quadratic phonon Hamiltonian $\mH^{(2)}$ in \refe{H2_diagonal}. 

The spatial correlator $\ev{\delta R_{j,s,\md} \delta R_{j',s',\md'}}$ between ions in unit cells $j$ and $j'$, sublattices $s$ and $s'$ and dimensions $\md$ and $\md'$ is given by 
\begin{widetext}
\spl{
\label{EQ:deltaR_deltaR}
\ev{\delta R_{j,s,\md} \, \delta R_{j',s',\md'}} = \frac{1}{ m_I \NI \sqrt{ \Omega_{\md} \Omega_{\md'}  }} \sum_{k,k'}\left[  e^{-id(kj-k'j')}  \ev{ {a^\dag_{k,s,\md}}   {a_{k',s',\md'}} } + e^{id(kj-k'j')}  \ev{ {a_{k,s,\md}}   {a^\dag_{k',s',\md'}} } \right. \\
\left. + e^{-id(kj+k'j')}  \ev{ {a^\dag _{k,s,\md}}  {a^\dag_{k',s',\md'}}  } + e^{id(kj+k'j')}  \ev{ {a_{k,s,\md}}   {a_{k',s',\md'}} }  \right].
}
The four correlators of the type $ \ev{ {a^\dag _{k,s,\md}}  {a^\dag_{k',s',\md'}}  }, \;\ev{ {a_{k,s,\md}}   {a_{k',s',\md'}} }, $ etc., appear ubiquitously in the evaluation of observables composed of products of two position or momentum operators (analogous to single-particle operators in the formalism of second quantization). These can subsequently be evaluated and expressed explicitly as functions of the elements of $\bx, \bp, \bq$ and $\omega_{k,\gamma}$
\al{
\ev{ {a^\dag_{k,s,\md}}   {a_{k',s',\md'}} } =& 
\sum_{\gamma,\gamma'} \left[ 
{u_{s,\md}^{(k,\gamma)}}^* {u_{s',\md'}^{(k',\gamma')}}  \ev{ \beta^\dag_{k,\gamma}   \beta_{k',\gamma'} }
+ {v_{s,\md}^{(k,\gamma)}} {v_{s',\md'}^{(k,\gamma)}}^* (\ev{ \beta_{k',\gamma'}^\dag   \beta_{k,\gamma} } + \delta_{k,k'} \delta_{\gamma,\gamma'} )   \right] \nonumber\\&
+\delta_{k,0} \, \delta_{k',0}  \sum_{n  } \left[  {v_{s,\md}^{(0,n)}}^* {v_{s',\md'}^{(0,n)}} \ev{\mathscr P_n^2}
+ {u_{s,\md}^{(0,n)}}^* {u_{s',\md'}^{(0,n)}} \ev{\mathcal Q_n^2} \right]  \label{EQ:akd_ak}\\
\ev{ {a^\dag_{k,s,\md}}   {a^\dag_{k',s',\md'}} } =& 
-\sum_{\gamma,\gamma'} \left[ 
{u_{s,\md}^{(k,\gamma)}}^* {v_{s',\md'}^{(k',\gamma')}}  \ev{ \beta^\dag_{k,\gamma}   \beta_{k',\gamma'} }
+ {v_{s,\md}^{(k,\gamma)}} {u_{s',\md'}^{(k,\gamma)}}^* (\ev{ \beta_{k',\gamma'}^\dag   \beta_{k,\gamma} } + \delta_{k,k'} \delta_{\gamma,\gamma'} )   \right] \nonumber \\&
+ \delta_{k,0} \, \delta_{k',0}  \sum_{n } \left[ {v_{s,\md}^{(0,n)}}^* {v_{s',\md'}^{(0,n)}}^* \ev{\mathscr P_n^2}
- {u_{s,\md}^{(0,n)}}^* {u_{s',\md'}^{(0,n)}}^* \ev{\mathcal Q_n^2} \right] \label{EQ:akd_akd}.
}
\end{widetext}

\pic{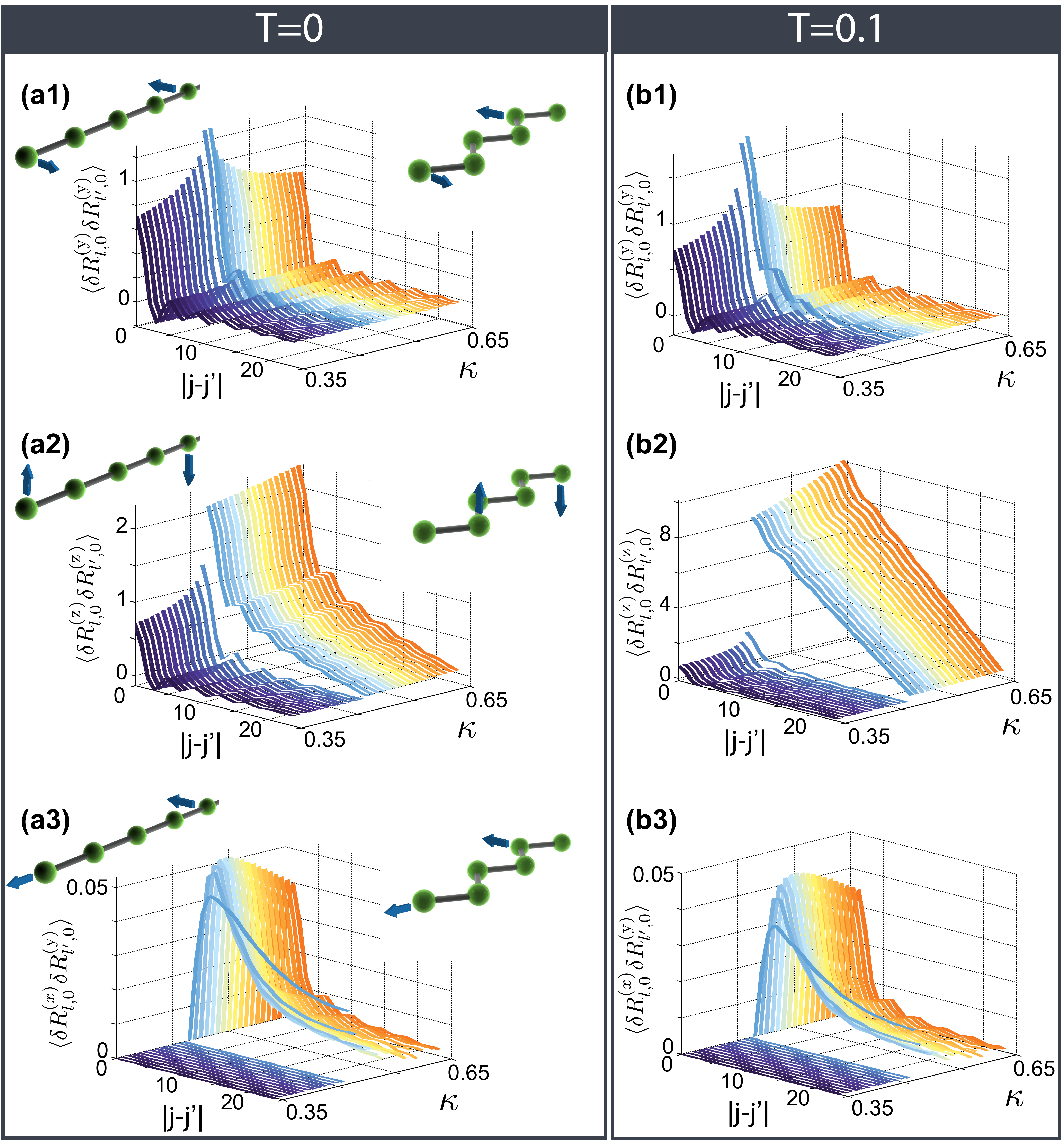}{\label{FIG:spatial_correlators} Spatial correlations $\ev{\delta R_{j,s,\md} \delta R_{j',s',\md'}}$ as a function of the distance $j-j'$ in an infinite ion chain at $T=0$ (left column) and finite temperature $T=0.1 \omega_I^{-1}$ (right column). These are shown for ions within the same sublattice $s=s'$ and as a function of the radial trapping frequency $\omega$ (directly related to $\kappa$) across the transition at $\omega_c$. For $\omega >\omega_c$ the chain is in a linear configuration while for $\omega <\omega_c$ in the zigzag configuration. Panels (a1, b1) and (a2, b2) show the transverse correlations $\ev{\delta R_{j,s,y} \delta R_{j',s,y}}$ and $\ev{\delta R_{j,s,z} \delta R_{j',s,z}}$ within and out of the zigzag plane respectively. On the linear side of the transition at $\kappa<\kappa_c$, these are all equal by virtue of the rotational symmetry around $\be_x$. Panels (a3,b3) depict the cross-dimensions correlation $\ev{\delta R_{j,s,x} \delta R_{j',s,z}}$ (shown here without the constant offset $\ev{\delta R_{j,s,\md} \delta R_{j',s',\md'}}_\mQ$), which is only non-zero in the zigzag configuration and the spontaneous breaking of symmetry couples the bare motion along these directions2. Note that $j$ does not refer to the ion index, but is the dimerized unit cell's index, while $s=s'$. The ion chain drawings in each panel of the left column depict the chain configuration and the correlation considered. For (a2, b2) the correlator is calculated for a finite system of $\NI=50$ ions, whereas the other correlators converge in the thermodynamic limit, for which they are shown.}{\linewidth}

Here we introduced the additional summation index $n$, which runs over all effective free particle degrees of freedom: $n \in \{ \mbox{l,r} \}$ (longitudinal or radial) in the zigzag or $n \in \{ \mbox{l} \}$ in the linear regime. At $T=0$ only the phonon ground state contributes, whereas at non-zero $T$, the population of the mode $(k,\gamma)$ is given by the Bose-Einstein factor
\al{
 \ev{\beta_{k,\gamma}^\dag \beta_{k',\gamma'} } &= \delta_{k,k'} \delta_{\gamma,\gamma'}  \frac{1}{e^{\omega_{k,\gamma / T}}-1}\\
\ev{\beta_{k,\gamma}\beta_{k',\gamma'} } &= \ev{\beta_{k,\gamma}^\dag \beta_{k',\gamma'}^\dag }=0
}
All cross terms $(k,\gamma) \neq (k',\gamma')$ vanish, since the density matrix is a statistical mixture of eigenstates of phonon Fock states. We point out that the phonon-related terms in \refe{akd_ak} and \refe{akd_akd} are independent of the eigenvectors' overall complex phase. This is consistent with the fact that a physical observable is independent of the arbitrary phase. In contrast, the $\ev{\mathcal Q^2}$ terms explicitly depend on the overall phase of the vector $\bp$. This may seem inconsistent at first, however, unlike for the phonon eigenvectors, the global phase of these vectors is uniquely fixed by the additional constraints discussed in \refs{properties_zero_subspace}.

By first replacing $\delta R_{j,s,\md} \propto b_{j,s,\md}+b_{j,s,\md}^\dag$ in both position operators in \refe{deltaR_deltaR}, one notices that the $\mathscr P$ contribution vanishes and, consequently, all terms involving $\mathscr P$ in Eqs.~(\ref{EQ:akd_ak}, \ref{EQ:akd_akd}) also cancel. On the other hand,  let us evaluate the terms containing a $\ev{\mathcal Q^2}$ in Eqs.~(\ref{EQ:akd_ak}, \ref{EQ:akd_akd}). For the helical case, given the form of the vector $\bq=\sqrt{N}[0,0,0,0,-i,i,0,0,0,0,-i,i]^t/2$ and \refe{ev_Qsq}, this contribution becomes
\spl{
\ev{\delta R_{j,s,\md} \, \delta R_{j',s',\md'}}_\mQ&= (-1)^{s-s'} \frac{1}{2\mI \Omega_z}(u_{j,s,\md}^{(0)} -u_{j,s,\md}^{(0)^*} )\\
&\quad \times  (u_{j',s',\md'}^{(0)} -u_{j',s',\md'}^{(0)^*} ) \ev{\mQ^2}\\
&=(-1)^{s-s'} \frac{\pi^2 \lambda^2 \Delta_0^2}{3 \mI \omega_I}
\label{EQ:deltaR_deltaR_Q_contrib}
}
in addition to the phonon contribution. In the axial case, the expression is analogous with $\NI d$ replacing $\Delta_0$ and the factor $(-1)^{s-s'}$ falling away. The contributions from the zero energy degrees of freedom to the spatial correlators given in \refe{deltaR_deltaR_Q_contrib} is thus simply a constant (distance-independent, but sublattice-dependent) offset in the respective dimension. For the helical case, this contribution is finite (since $\ev{\mQ^2}$ is intensive), whereas for the axial case, the offset diverges (here $\ev{\mQ^2} \propto \NI^2$, as the circumference scales linearly with $\NI$) in the thermodynamic limit $\NI \to \infty$. This divergence is analogous and of similar origin as the divergence of the local spatial variance discussed in \refs{Ginzburg_crit}.

Let us shortly discuss the experimental relevance of these terms, which are relevant if the system is in a pure eigenstate of the effective free particle sector of $\mH$, where all ions are collectively delocalized along the respective direction with PBCs. Two effects limit this in experiments: 1) The geometry for 1D ion strings is generally not ring-shaped, but rather a confined 1D string. Although the PBCs are required to characterize the phonons using a quasi-momentum, the zero energy mode will not appear in this direction if the entire ion string is axially trapped. We do not include this diverging offset contribution in the axial correlators shown in \reff{spatial_correlators}. 2) The strong coupling of the charged ion to electric fields makes the delocalized eigenstates susceptible to decoherence. For the emerging effective helical zero energy degree of freedom, the severity of the decoherence should increase with $\Delta_0$. The most promising regimes to observe these effects should thus be close to the zigzag transition, at small $\Delta_0$.

In \reff{spatial_correlators} we show such correlators in real-space as a function of $\kappa$, as one crosses the zigzag transition. The left column shows results for the zero temperature case, while the right column describes the finite-temperature case. In each panel the same range of $\kappa$ is shown, crossing from the linear to the zigzag phase from left to right. We find that an increase in temperature (right column in \reff{spatial_correlators}) generally increases the spatial correlations, as both thermal and quantum fluctuations contribute. The infrared divergence, which forbids the formation of true long-range order (e.g. phase rigidity of the zigzag structure), leads to a constant, distance-independent offset on the length scales shown in \reff{spatial_correlators},  as the divergence originates from a contribution at small $k$. This contribution is in addition to the constant contribution from the effective free particle degree of freedom in \refe{deltaR_deltaR_Q_contrib}.

\textit{Transverse correlations:} In the linear regime the transverse correlations decay exponentially, initially oscillating around zero. As expected from symmetry, the transverse correlator for $\nu=\nu=y$ shown Fig.~\ref{FIG:spatial_correlators} (a1, b1) and for $\nu=\nu'=z$ in Fig.~\ref{FIG:spatial_correlators} (a2, b2) directions are identical for $\alpha=1$. This changes drastically at the transition point, where the zigzag pattern forms in the $x$-$y$-plane: while the $y$-$y$ correlator at fixed distance decays on both sides of the transition with $|\kappa-\kappa_c|$, the $z$-$z$ correlator becomes large on the zigzag side, increasing for larger $\kappa > \kappa_c$. This can be understood from the emergence of the additional soft helical mode, the virtual occupation of which in the phonon ground state $\ket{\psi_0}$ is larger than for other modes, as these excitations have a low energy penalization. Both $\ev{\delta R_{j,s,x} \delta R_{j',s,z} }$ and $\ev{\delta R_{j,s,y} \delta R_{j',s,z} }$ vanish on each side of the transition, both for $0$ and finite temperature. However, as shown in Fig.~\ref{FIG:spatial_correlators} (a3, b3), the $x-y$ correlator is non-zero in the zigzag regime.

\section{Heat Capacity} \label{SEC:heat_capacity}
Having determined all the eigenmodes and energies, we compute the heat capacity of the system, which is given by 
\spl{
C&=\pd{\ev \mH}{T}\\
&=\frac{1}{T^2} \pd{^2}{\beta^2} \ln Z\\
&=\frac{1}{T^2}\sum_{k,\gamma} \frac{\omega_{k,\gamma} e^{-\omega_{k,\gamma}/T } }{ (  e^{-\omega_{k,\gamma}/T } -1)^2 }
}
where $T$ is the temperature. 
\pich{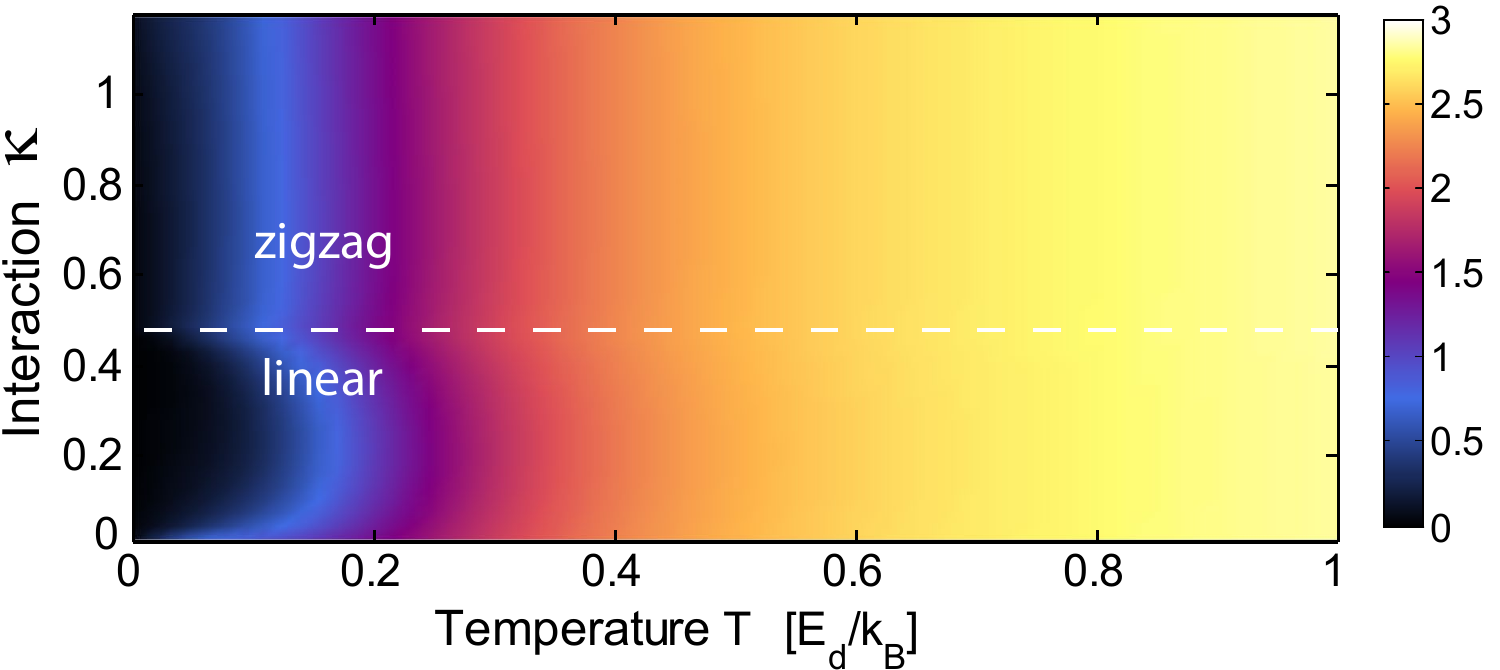}{\label{FIG:heat_capacity_of_omegaI} Specific heat capacity $c=C/\NI$ as a function of temperature across the zigzag transition shown in the thermodynamic limit $\NI\to \infty$. In the zigzag regime the additional low energy modes lead to an enhancement of the heat capacity relative to the linear regime.}{0.9\linewidth}
This is shown in \reff{heat_capacity_of_omegaI} as a function of $\kappa$ and the temperature. 
At low temperatures $T$, one finds a pronounced lobe of lower heat capacity than in the zigzag regime. This can be understood from the absence of the additional helical twisting low energy mode in the linear regime, which reduces the density of states at low energies and thus also the heat capacity. In the high temperature limit, the Dulong-Petit law $c=C/ \NI = 3$ is recovered for all $\kappa$. Both contributions from the phonon modes, as well as the effective free particle degrees of freedom have to be taken into account when evaluating the heat capacity. The relative weight for the latter however vanishes in the specific heat capacity $c$ as $\NI \to \infty$.

\section{Local Susceptibility} \label{SEC:local_susceptibility} 
Since in ion experiments it is common practice to illuminate a single ion in the chain with a laser, we calculate the local spatial susceptibility in the limit of a large chain, where the phonon modes form continuous bands. The spatial susceptibility relates the local displacement $\ev{\delta \bR_{j,\md} }$ of an ion $j$ to a temporally periodic driving force,  described by a time-dependent potential $\propto \cos(\omega t) \delta \bR_{j,\md}$. With both the observable and the perturbing potential being the position operator of one ion, the local susceptibility (in linear response) is given by \cite{Chaikin1995}
\spl{
\label{EQ:susceptilibility_freq}
\chi_{(j,s,\md)(j,s,\md)}(\omega)=&\frac{1 }{Z} \sum_{m,m'} \frac{ \bra{E_m} \delta \bR_{j,\md} \ket{E_{m'}} \bra{E_{m'}} \delta \bR_{j,\md} \ket{E_{m}}   }{\omega-(E_{m'}-E_{m}) + i 0^+ }\\  &\times [ e^{-E_m/T } - e^{- E_{m'}/T } ].
}
Here $\ket{E_m}$ are the many-body eigenstates, which to a good approximation (since higher order terms are small) are phonon number states specified by a set of integer occupation numbers $\{ n_{k,\gamma} \}$ for each phonon mode
\spl{
\ket{E_m}=\ket{ \{ n_{k,\gamma}  \}  }=\prod_{k,\gamma} \frac{1}{\sqrt{ n_{k,\gamma}! }}(\beta_{k,\gamma}^\dag)^{n_{k,\gamma}} \ket{\psi_0}.
}
\pich{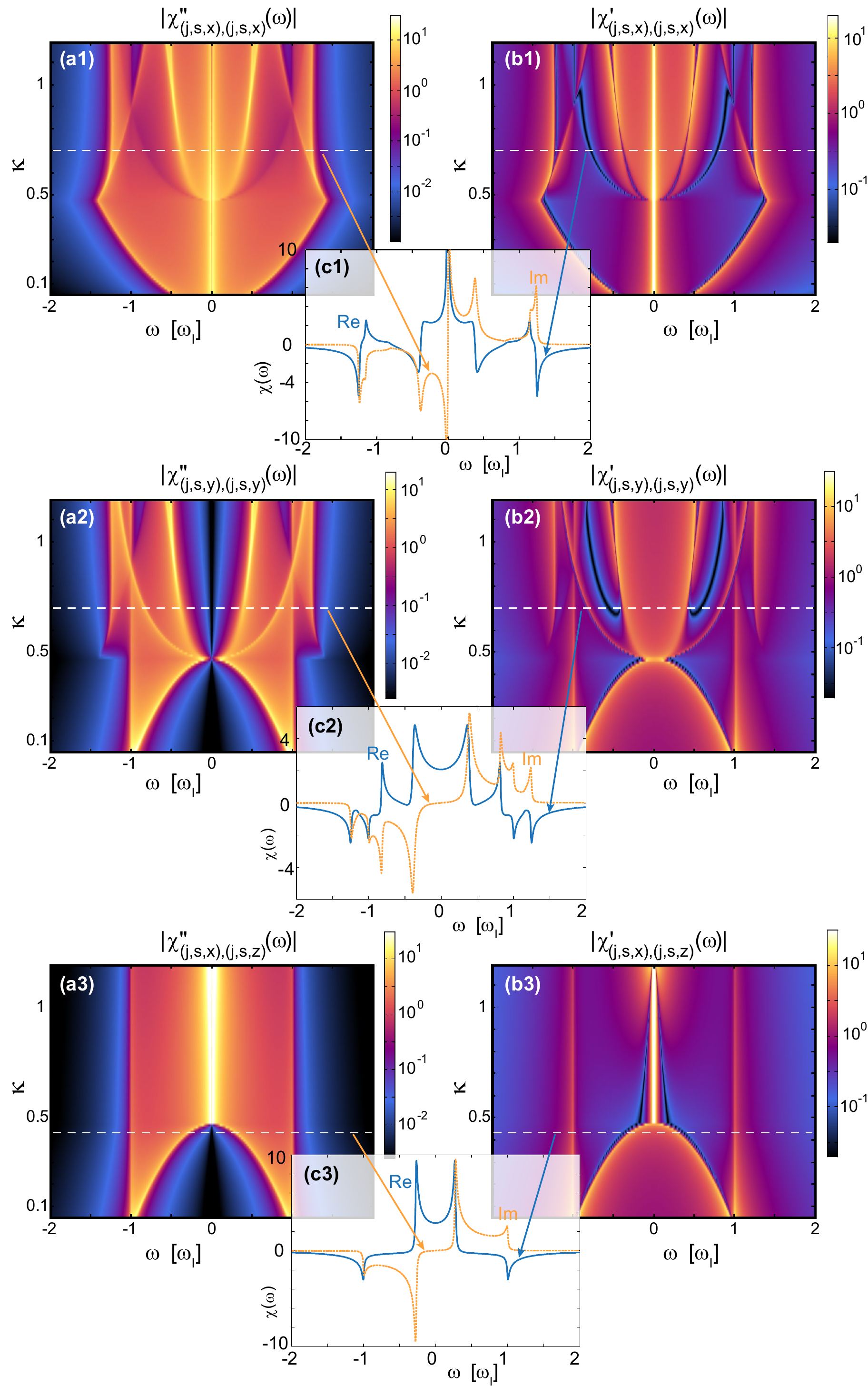}{\label{FIG:chi_of_kappa_combined} The local dynamical susceptibility $\chi_{(j,s,\md),(j,s,\md)}(\omega)$ in units of $d^2/\omega_I$ at $T=0$ as a function of frequency $\omega$ and the dimensionless interaction strength $\kappa$. The panels on the left (a1-a3) show the imaginary, dissipative part of $\chi$ (using a logarithmic color scale), whereas the panels on the right (b1-b3) show the real, reactive part. The three rows are the different diagonal components, i.e. the longitudinal $\chi_{(j,s,x),(j,s,x)}(\omega)$ and transverse components $\chi_{(j,s,y),(j,s,y)}(\omega)$, $\chi_{(j,s,z),(j,s,z)}(\omega)$ components for rows one to three respectively. The insets (c1-c3) show the susceptibility on a linear scale along the slices indicated as dashed white lines, located at $\kappa=0.7$, $\kappa=0.7$ and $\kappa=0.44$ respectively.  A broadening parameter of $\eta=10^{-2}$ was chosen to broaden the spectral peaks and achieve continuous functions, corresponding to a large system in the thermodynamic limit.
}{\linewidth}
The local dynamical susceptibility is a tensor consisting of $6 \times 6$ components. We focus on the diagonal components, i.e. for the driving force and the observed response pointing in the same three natural directions of the system. Due to the symmetry of the unit cell, these are pairwise equal and in \reff{chi_of_kappa_combined} we show the three independent $\chi_{(j,\md,s),(j,\md,s)}(\omega)$ (corresponding to different spatial orientations) as a function of frequency and interaction strength $\kappa$ at zero temperature. Since the local ion position operator $\delta \bR_{j,\md}$ decomposes into a single phonon creation and annihilation operator, only single phonon states and the phonon ground state contribute in the evaluation of Eq.~(\ref{EQ:susceptilibility_freq}) at $T=0$. 

This leads to the expression 
\spl{
\chi_{(j,s,\md),(j,s,\md)}(\omega)=& - \frac{1}{\mI \Omega_{\md} \NI} \sum_{k,\gamma}\left[ \frac{|u_{s,\md}^{(k,\gamma)}+v_{s,\md}^{(k,\gamma)}|^2}{\omega + \omega_{k,\gamma}  + i \eta} \right. \\  & \left. - \frac{|u_{s,\md}^{(k,\gamma)}+v_{s,\md}^{(k,\gamma)}|^2}{\omega - \omega_{k,\gamma} + i \eta}  \right],
}
which is shown as a function of $\kappa$ (the effective ion coupling strength) and $\omega$ in \reff{chi_of_kappa_combined}.

 In the linear regime, the two radial susceptibilities agree, since the $O(2)$ symmetry has not been broken yet. In the zigzag regime, however, the system responds very differently to a driving force along these directions with the appearance of new modes and van Hove singularities leading to pronounced peaks in $\chi(\omega)$. The closing of the radial excitation gap and opening and new gapped modes can clearly be seen across the transition. In addition, as the high lying modes get admixed after $\kappa > \kappa_c$, this is also reflected by the additional contribution in $\chi_{(j,y,s),(j,y,s)}(\omega)$, which \textsl{'smears in'}, as the admixture grows.

The imaginary part of the local susceptibility $\chi''_{jj}=\mathcal{I}\left(\chi_{jj}\right)$ indicates the amount of energy that the system absorbs to linear order in the driving amplitude when the system is driven at a frequency by $V(t)=A \, \cos(\omega t) \, \delta \bR_{j,s,\md}$. Specifically, the average energy absorption rate $\frac{d \overline E}{dt} = 2 A^2 \omega \, \chi''(\omega)$ is shown in \reff{chi_of_kappa_combined} (a1-a3).

The real part $\chi'_{jj}=\mbox{Re}\left(\chi_{jj}\right)$, shown in \reff{chi_of_kappa_combined} (b1-b3), describes the reactive part, i.e. response of the position of the ion to lowest order in the limit of weak perturbation.

The complex angle of the susceptibility, ${\rm arg}(\chi)$, indicates the phase shift of a given ion's motion relative to the local driving force and is shown in \reff{chi_angle_of_kappa_combined}. Essentially, the phase shift is zero (or $\pi$) if the system is driven at a frequency which lies outside the phonon frequency band $\omega_{k,\gamma}$ of any mode coupling to the perturbing operator. Specifically in the radially gapped regime with radial driving, this phase shift is essentially zero for a large lobe-shaped region at low frequency, shown in \reff{chi_angle_of_kappa_combined} (b,c). Here, the ion locally moves in phase with the driving force, whereas when driving above the phonon bands leads to a phase difference of $\pm \pi$. In both cases, no energy is absorbed to linear order, as is to be expected when not driving at a collective resonance frequency. In the intermediate region where $\omega$ reaches certain phonon frequencies and the phonon structure changes with $\kappa$, also changing the coupling, an intricate structure is also observed for the local phase shifts.

\pich{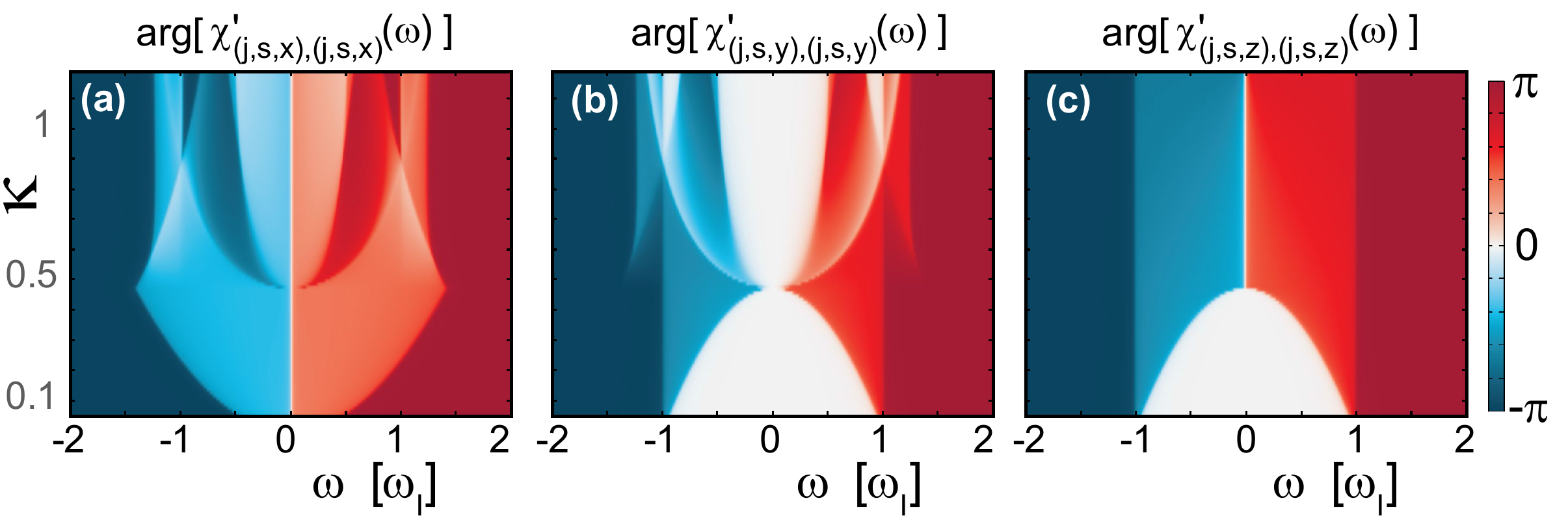}{\label{FIG:chi_angle_of_kappa_combined} The complex angle of the susceptibility, ${\rm arg}(\chi_{(j,s,\md),(j,s,md)}(\omega))$, corresponding to the phase shift between the observable and the driving force, for different values of the frequency $\omega$ and the dimensionless parameter $\kappa$. Panels (a), (b), (c) show this phase shift for the different diagonal components $\md \in \{ x,y,z \}$ respectively.}{\linewidth} 

\section{Correlation Energy Reduction}
\label{SEC:correlation_energy_reduc}
Another quantity of interest is the correlation energy reduction 
\spl{
\Delta E_0 = \bra{\psi_0} \mHph \ket{\psi_0} - \bra{\psi_{\mbox{\tiny 0,uncorr}}} \mHph \ket{\psi_{\mbox{\tiny 0,uncorr}}} 
}
of the phonon ground state $\ket{\psi_0}$ relative to the general variational uncorrelated state $\ket{\psi_{\mbox{\tiny 0,uncorr}}}= \prod_{\otimes l} \ket{\varphi_0}_l$. The latter is optimized over all states to minimize $\bra{\psi_{\mbox{\tiny 0,uncorr}}} \mHph \ket{\psi_{\mbox{\tiny 0,uncorr}}}$. 
Thus $\Delta E_0 $ is the amount of energy by which the system can reduce its ground state energy via the generation of correlations and entanglement between the different ions, which we show in \reff{energy_reduction_gs} as a function of $\kappa$. Note that $\Delta E_0$ is negative, as the phonon ground state energy is lower than that of $\ket{\psi_{\mbox{\tiny 0,uncorr}}}$.

\pich{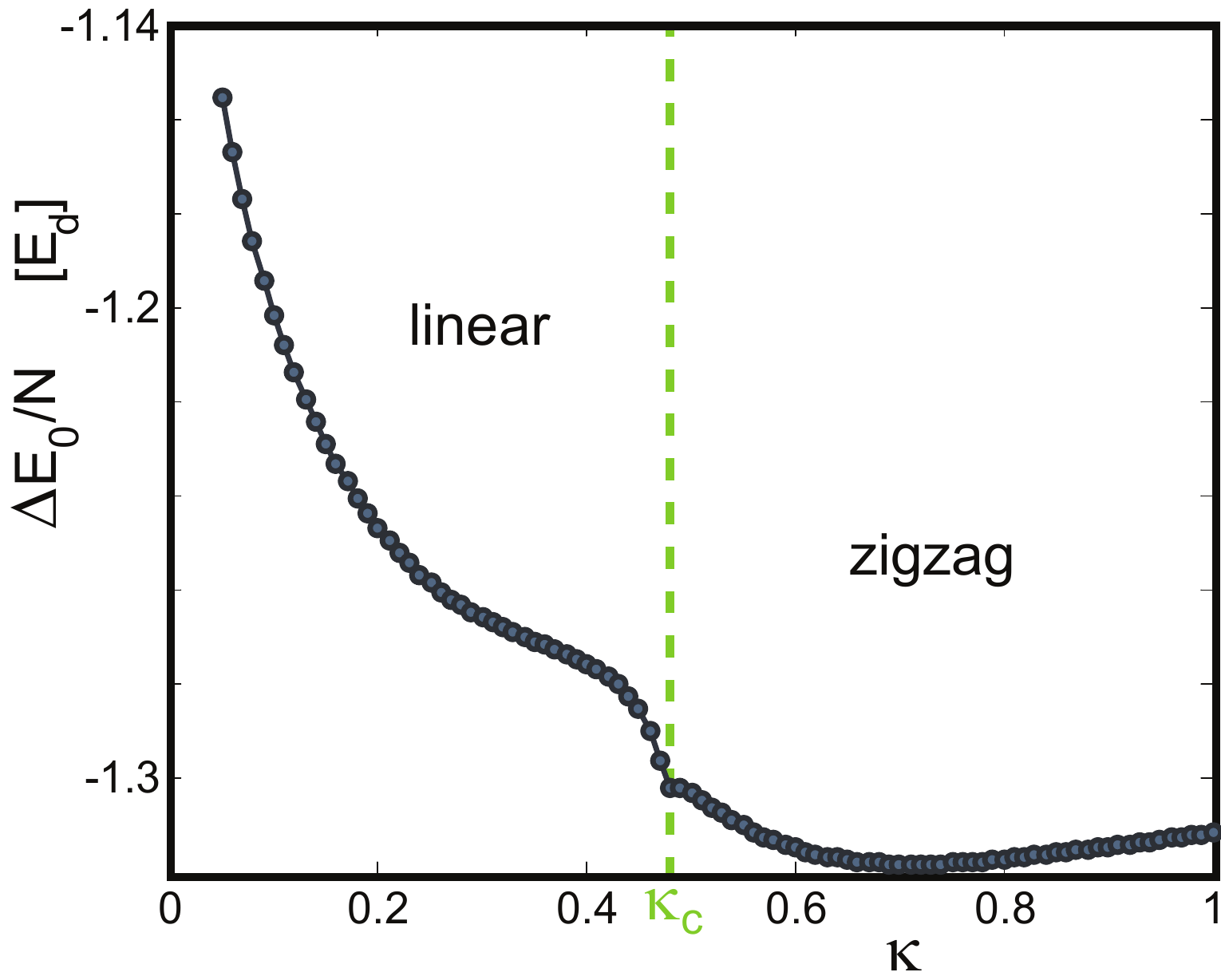}{\label{FIG:energy_reduction_gs} Correlation energy reduction per ion $\Delta E_0 / N. \;$ $\Delta E_0$ is the amount by which the system lowers its energy in the ground state $\ket{\psi_0}$ relative to the uncorrelated local product state of local ion ground states $\ket{\psi_{\mbox{\tiny 0,uncorr}}}$, by introducing quantum fluctuations. This is shown as a function of $\kappa$ in the limit of large $N$, where $\Delta E_0 / N$ is an intensive function, although the total energy per ion at fixed $d$ is a super-extensive function.}{7cm}
Since the local potential of any ion is well approximated by a quadratic potential (except for a small region $\kappa \approx \kappa_c$ ) $\bra{\psi_{\mbox{\tiny 0,uncorr}}} \mHph \ket{\psi_{\mbox{\tiny 0,uncorr}}}$ is very close to the minimum variational energy with respect to the set of all uncorrelated ion product states. For small $\kappa$ in the linear regime, $\Delta E_0$ decreases with increasing $\kappa$, as the interactions coupling terms grow and the system can reduce its energy by generating entanglement. At the zigzag transition $\kappa=\kappa_c$, the correlation energy reduction features a kink (as would be expected from the energy being a continuous function of $\kappa$). In the zigzag regime, ${\Delta E_0}/{E_d}$  shows a much weaker dependence on $\kappa$ and even starts increasing again, as the ions can avoid each other by increasing the zigzag parameter $\Delta_0$.

\section{Conclusions} \label{SEC:conclusions}   
In conclusion, we have presented an alternative approach to calculate, from first principles, the phonon spectrum and operators of a trapped ion chain, emphasizing on the close analogy to Bogoliubov theory. In particular, we show that the diagonalization of the quadratic system is not sufficient to treat all degrees of freedom in the presence of spontaneously broken symmetries. Here, we presented how the most general normal form of the quadratic Hamiltonian can be constructed, which contains a non-bosonic, effective free particle degree of freedom for each broken symmetry. Concentrating on systems with radially symmetric trapping ($O(2)$-symmetric), we then use the formalism to analyze in detail linear structures like a linear chain and a zigzag chain and study the behavior of the mode structure across the transition. To demonstrate how the formalism can be applied to evaluate physical observables, we calculate the spatial correlations, heat capacity, dynamical susceptibility and the correlation energy reduction. The latter underlines the collective nature of the phonon ground state, which cannot be understood in terms of individual local ion ground states.
The formalism presented in this work provides a convenient platform to study a variety of time-independent and dynamical phenomena. For instance, an application to study the effect of non-linear terms \cite{Porras2004,Deng2008,Dutta2013,Dutta2016}, magnetic fields on the phonon structure and observables \cite{Piacente2003,Bermudez2011,Bermudez2015}, or the influence of micromotion \cite{Nguyen2012} and the dynamics it incurs, would be of interest.

{\bf Acknowledgments:} We thank R. Gerritsma for fruitful discussions. This work is supported by the Singapore MOE Academic Research Fund Tier-2 project (Project No. MOE2014-T2-2-119, with WBS No. R-144-000-350-112) (U.B. and D.P.).  W.H. acknowledges financial support from the Deutsche Forschungsgemeinschaft (DFG) via Sonderforschungsbereich SFB/TR 49.


\clearpage
\begin{appendix}

\section{Expansion Terms for the Linear Chain}\label{APP:taylor_exp_ion_pot}     
For the linear chain, the Coulomb interaction potential can be written as 
\spl{
V(\dx, \dy,\dz)&=\sum_j \frac{m_I \omega_I^2}{2} (\dy_j^2 + \dz_j^2) \\& + \frac{q^2}{4 \pi \epsilon_0} \sum_{l<j} \frac{1}{\sqrt{ g_{l,j}(\delta \br_l, \delta \br_j)}},
}
where
\spl{
 g_{l,j}(\delta \br_l, \delta \br_j)&= [(l-j)d + \dx_l-\dx_j]^2  \\& + [\dy_l-\dy_j]^2+[\dz_l-\dz_j]^2
}
is a function of the coordinates of ions $l$ and $j$ only.
The first derivatives of the potential with respect to the individual coordinates are
\al{
\left. \pd{V}{\dx_l} \right|_{\bdelta\br_j=0} &= -\frac{q^2}{4 \pi \epsilon_0} \sum_{j\neq l} \frac{(l-j)d}{|(l-j)d|^3}
\left. \pd{V}{\dy_l} \right|_{\bdelta\br_j=0} \\&=m_I \omega_I^2 \dy_l +  -\frac{q^2}{4 \pi \epsilon_0} \sum_{j\neq l} \frac{(l-j)d}{|(l-j)d|^3}.
}

\section{Evaluation of $f_{m,m'}$}
\label{APP:vanishing_off_diagonal} 
We restrict the proof to the case of even $\NI$ and PBCs. The thermodynamic limit $\NI \to \infty$ at fixed inter-ion spacing $d$ can be drawn subsequently. The aim of this proof is to show that the coupling matrix elements $f_{m,m'}$ are only non-vanishing if $m=m'$. 

Inserting Eq.~(\ref{EQ:quasi-momentum_parametrization}) into Eq.~(\ref{EQ:def_f_m_md}), one has
\spl{
\label{EQ:def_f_m_md_inserted}
f_{m,m'} = \frac{1}{N} \sum_{l,l'=0}^{N-1}  e^{i \pi (l-l' )} e^{-\frac{2 \pi i}{N} (ml -m' l')} f(\Dis(l-l')).
}
Performing a transformation of summation variables by defining
\al{
\label{EQ:variable_trafor_pL}
p &= l-l'\\
L &= l+l'
}
and using the property
\spl{
ml -m'l' &= \frac 1 2 (m+m')p + \frac 1 2 (m-m')L
}
the sum can be split into three parts
\spl{
\label{EQ:def_f_m_md_split}
f_{m,m'} &= \frac{1}{N} \sum_{l=0}^{N-1}  e^{-\frac{2 \pi i}{N} (m -m' )l} f(\Dis(0)) \\
&+ \frac{1}{N} \sum_{l=1}^{N-1} \sum_{l'=0}^{l-1}  e^{i \pi (l-l' )} e^{-\frac{2 \pi i}{N} (ml -m' l')} f(\Dis(l-l'))\\
&+ \frac{1}{N} \sum_{l=0}^{N-2} \sum_{l'=l+1}^{N-1}  e^{i \pi (l-l' )} e^{-\frac{2 \pi i}{N} (ml -m' l')} f(\Dis(l-l')).
}
The diagonal sum in the first line of Eq.~(\ref{EQ:def_f_m_md_split}) vanish, since for our case $f(0)=0$. For the sums in the second and third lines we change the summation procedure using \refe{variable_trafor_pL}, now working in terms of variables $p$ and $L$. This also requires the change of the summation range
\al{
\sum_{l=1}^{N-1} \sum_{l'=0}^{l-1} \ldots & \mapsto \sum_{p=1}^{N-1} \sum_{\stackrel{L=p}{\mbox{\tiny step $2$}} }^{2(N-1)-p } \ldots\\
\sum_{l=0}^{N-2} \sum_{l'=l+1}^{N-1} \ldots & \mapsto \sum_{p=1- N}^{-1} \sum_{\stackrel{L=-p}{\mbox{\tiny step $2$}} }^{2(N-1)+p } \ldots
}
and hence
\spl{
\label{EQ:def_f_m_md_split_eval1}
f_{m,m'} &= 
\sum_{p=1}^{N-1} (-1)^{p}  e^{-i\frac{\pi }{N} (m +m' ) p} f(\Dis(p)) \\
&\times \frac{1}{N} \sum_{\stackrel{L=p}{\mbox{\tiny step $2$}} }^{2(N-1)-p }    e^{-i\frac{\pi}{N} (m -m' ) L}\\
&+\sum_{p=1- N}^{-1}  (-1)^{p}  e^{-i\frac{\pi }{N} (m +m' ) p} f(\Dis(p)) \\
&\times \frac{1}{N} \sum_{\stackrel{L=-p}{\mbox{\tiny step $2$}} }^{2(N-1)+ p }    e^{-i\frac{\pi}{N} (m -m' ) L}.
}
Now, in the lowest two lines of Eq.~(\ref{EQ:def_f_m_md_split_eval1}), we perform a change of summation variables $p \mapsto - p$, such that the last sums of both terms become identical. For the case $m\neq m'$ the two summations over the exponential factors not containing can be evaluated subtracting the expressions for the geometric series 
\spl{
\sum_{k=0}^{n-1} r^k = \frac{1-r^n}{1-r}
}
and one obtains
\spl{
 &\sum_{\stackrel{L=p}{\mbox{\tiny step $2$}} }^{2(N-1)-p }    e^{-i\frac{\pi}{N} (m -m' ) L} =\frac{ e^{-i \pi \frac{m-m'}{N} p  }  - e^{i \pi \frac{m-m'}{N} p }  }{ 1- e^{-2 \pi i \frac{m-m'}{N}  } }.
}
Inserting this into Eq.~(\ref{EQ:def_f_m_md_split_eval1}) and multiplying out the terms, we find
\spl{
\label{EQ:def_f_m_md_split_eval2}
f_{m,m'} &= \frac{1}{ 1- e^{-2 \pi i \frac{m-m'}{N}  } } 
\frac{1}{N}
\sum_{p=1}^{N-1}  (-1)^{p}  f(\Dis(p))\\
&\times  \Big[ 
  e^{-2 \pi i\frac{m}{N} p} 
-e^{-2 \pi i\frac{m'}{N} p} 
+  e^{2 \pi i\frac{m'}{N} p} 
-e^{2 \pi i\frac{m}{N} p} 
\Big].
}
In the last two exponential terms in brackets, we now perform a summation variable transformation $\mbox{$p \mapsto \tilde{p}=N-p$}$. We then have $\Dis(\tilde{p})=\Dis(p)$ and, if $N$ is even,  $(-1)^{\tilde{p}}=(-1)^{{p}} $. Inserting these relations into Eq.~(\ref{EQ:def_f_m_md_split_eval2}) and renaming $\tilde{p}$ to ${p}$, one finds that all exponential factors cancel pairwise and we have
\spl{
\label{EQ:def_f_m_md_split_eval3}
f_{m,m'}=0   \qquad \mbox{if $m\neq m'$}.
}

\section{Position and conjugate momentum operator for a free 1D particle with Born-von Karman PBCs}
\label{SEC:PBC_position_op}

We seek to construct an effective position operator $\mQ$ fully accounting for the PBCs, which is conjugate to $\mathscr P$ in the sense
\al{
e^{i\Delta p \mQ } \mathscr P e^{-i\Delta p \mQ } &=  \mathscr P + \Delta p \\
e^{-i\Delta x \mathscr P } \mQ e^{i \Delta x \mathscr P } &= \lfloor \mQ + \Delta x \rfloor
}
where $\lfloor  \ldots \rfloor$ denotes the folding back of any eigenvalue shifted outside the range of defined position (angle) values into the main sector, i.e. modulo PBCs.

Let us work in the basis of momentum eigenstates $\ket l$ and construct $\mQ$ on a truncated, finite space. Specifically, let us work in the $(2M+1)$-dimensional subspace $\mH_M$ of states $l \leq M$. Within this subspace, we may ask what is the most localized (in $\varphi$) basis of states we can construct from the given basis set $\{ \ket l \}$? This basis of maximally localized angular states is given by 
\spl{
\ket{\phi_n}= \frac 1 {\sqrt{2M+1}} \sum_{l=-M}^M e^{-i \phi_n l} \ket l
}
and $\phi_n=\frac{2 \pi n}{2M+1}$, $n\in \{-M,\ldots, M\}$. Essentially, $\ket{\phi_n}$ are the maximally localized Wannier functions without a lattice, with the number of artificial lattice sites being $2M+1$. Being the unitary Fourier transform of an orthonormal basis $\ket l$, the angle states $\ket{\phi_n}$ constitute an orthonormal basis themselves $\braket{\phi_n}{\phi_{n'}}=\delta_{n,n'}$. The various $\phi_n(\varphi)=\braket{\varphi}{{\phi_n}}$ are identical up to translation, are centered at angular position $\phi_n$ and, in the limit of large $M$, each $\phi_n(\varphi)$ approaches a sinc-function.

We can now define a Hermitian angular position operator and the position operator $\mQ$ as a scaled version of it as
\spl{
\hat \varphi &:= \sum_{n=-M}^M \phi_n \ket{\phi_n}\bra{\phi_n}\\
\mQ&:= \frac{L}{2\pi} \hat \varphi,
}
both of which, by construction, have an equidistantly spaced spectrum on the periodic interval with the eigenstates $\ket{\phi_n}$. Here, $L$ is the circumference of the ring on which $\mQ$ acts. Importantly, it fully respects the PBCs in the sense that $\mQ$ is independent of the point where the artificial cut at the boundary is made and no effect of the boundary is is reflected in any observable.

It can be shown that 
\spl{
e^{i \mQ}= \left[ \sum_{l=-M}^{M-1} \ket{l+1}\bra{l} \right] + \ket{l=-M}\bra{l=M} 
}
which reflects the fact that $\mQ$ is the conjugate operator to $\mathscr P$, being the generator of discrete translation in $\mathscr P$ on the finite subspace $\mH_M$.

Having established that these are the appropriately extended operators of the ones obtained from the linear expansion of the ion chain, we can now proceed to evaluate the expectation values of interest. Primarily, one finds that in any eigenstate $\ket l$
\al{
\bra l \mQ \ket l &= 0\\
\bra l \mQ \mathscr P \ket l &= 0\\
\lim_{M \to \infty} \bra l \mQ^2  \ket l &= \frac{\pi^2 L^2}{3}.
}
The last expression agrees with the intuitive result of the spatial variance of a maximally delocalized state on a periodic ring.
\end{appendix}
\bibliography{ions}

\end{document}